\documentclass[preprint,3p,twocolumn]{elsarticle}
\biboptions{sort&compress}

\makeatletter
\def\ps@pprintTitle{%
	\let\@oddhead\@empty
	\let\@evenhead\@empty
	\let\@oddfoot\@empty
	\let\@evenfoot\@oddfoot
}
\makeatother

\usepackage[hyphens]{url}
\usepackage[hidelinks]{hyperref}

\usepackage{lineno}
\usepackage{optidef}
\usepackage{xcolor}
\usepackage{amsmath}
\usepackage{amssymb}
\usepackage{algpseudocode}
\usepackage{algorithm}
\usepackage{multirow}
\modulolinenumbers[5]
\journal{Applied Energy}
\usepackage{tikz}
\usepackage{pgfplots}
\usepackage{bm}
\pgfplotsset{width=7cm,compat=1.14}
\usepgfplotslibrary{statistics}
\usepackage{caption}
\usepackage{subcaption}
\usepackage[utf8]{inputenc}
\usetikzlibrary{arrows}
\usepackage{siunitx}
\usepgfplotslibrary{units}
\pgfplotsset{unit code/.code={\si{#1}}}
\usepackage{titlesec}

\usepackage{eurosym}
\DeclareSIUnit{\EUR}{\text{\euro}}

\newcommand{\specialcell}[2][c]{%
	\begin{tabular}[#1]{@{}c@{}}#2\end{tabular}}

\usepackage{colortbl}
\usepackage{xcolor}
\definecolor{green}{RGB}{173,255,47}%
\definecolor{red}{RGB}{255, 0, 0}%
\definecolor{orange}{RGB}{229, 83, 0}%
\definecolor{redPlots}{RGB}{204,37,41}%
\definecolor{greenPlots}{RGB}{62,150,81}%
\definecolor{bluePlots}{rgb}{0.00000,0.44700,0.74100}%
\definecolor{turqoise}{rgb}{0.00000,0.64700,0.64100}%

\newcommand{\mr}[1]{\mathrm{#1}}

\newcommand{\R}{\mathbb{R}}
\newcommand{\X}{\mathbf{X}}

\newcommand{\vc}[1]{\mathbf{#1}}
\newcommand{\mape}{\mathrm{sMAPE}}

\newcommand{\nin}{{n}}
\newcommand{\nout}{{m}}
\newcommand{\nhidn}{n}

\newcommand{\nts}{N}
\newcommand{\nruns}{T}

\newcommand{\nhyper}{z}
\newcommand{\hyperset}{Z}
\newcommand{\featset}{Z}
\newcommand{\nhypersub}{q}

\newcommand{\hid}{z}
\newcommand{\dt}{h}

\newcommand{\ixhid}{k}
\newcommand{\ixts}{k}

\newcommand{\ixset}{i}
\newcommand{\ixneuron}{i}

\newlength{\figW}
\newlength{\figH}
\newlength{\separ}

\titleformat{\paragraph}
{\normalsize\itshape}{\theparagraph}{1em}{}
\titlespacing*{\paragraph}
{0pt}{3.25ex plus 1ex minus .2ex}{1.5ex plus .2ex}
\usepackage[normalem]{ulem}

\usepackage[
type={CC},
modifier={by-nc-nd},
version={4.0},
]{doclicense}

\begin{document}

\begin{frontmatter}

\title{Forecasting day-ahead electricity prices in Europe: the importance of considering market integration}

\author[delft,vito]{Jesus Lago}
\ead{j.lagogarcia@tudelft.nl}

\author[vito]{Fjo De Ridder}
\author[vub]{Peter Vrancx}
\author[delft]{Bart De Schutter}
\address[delft]{Delft Center for Systems and Control, Delft University of Technology,\\ 
	Mekelweg 2, 2628CD Delft, The Netherlands}
\address[vito]{Energy Technology, VITO-Energyville,
	ThorPark, 3600 Genk, Belgium}
\address[vub]{AI Lab,
	Vrije Universiteit Brussel, Pleinlaan 2, 1050 Brussels, Belgium}
\cortext[cor]{Corresponding author}

\tnotetext[t1]{This is a preprint of the article: \textit{Forecasting day-ahead electricity prices in Europe: the importance of considering market integration, Applied Energy 211 (2018) 890-903}. \url{https://doi.org/10.1016/j.apenergy.2017.11.098}. This preprint is free of the publisher-introduced errors in the mathematical notation and the formating of the figures.}

\begin{abstract}
	Motivated by the increasing integration among electricity markets, in this paper we propose two different methods to incorporate market integration in electricity price forecasting and to improve the predictive performance. First, we propose a deep neural network that considers features from connected markets to improve the predictive accuracy in a local market. To measure the importance of these features, we propose a novel feature selection algorithm that, by using Bayesian optimization and functional analysis of variance, evaluates the effect of the features on the algorithm performance. In addition, using market integration, we propose a second model that, by simultaneously predicting prices from two markets, improves the forecasting accuracy even further. As a case study, we consider the electricity market in Belgium and the improvements in forecasting accuracy when using various French electricity features. We show that the two proposed models lead to improvements that are statistically significant. Particularly, due to market integration, the predictive accuracy is improved from 15.7\% to 12.5\% sMAPE (symmetric mean absolute percentage error). In addition, we show that the proposed feature selection algorithm is able to perform a correct assessment, i.e.~to discard the irrelevant features.
\end{abstract}

\end{frontmatter}

%\linenumbers

\section{Introduction}
As a result of the liberalization and deregulation of the electricity markets in the last two decades, the dynamics of electricity trade have been completely reshaped. In particular, electricity has become a commodity that displays a set of characteristics that are uncommon to other markets: a constant balance between production and consumption, load and generation that are influenced by external weather conditions, and dependence of the consumption on the hour of the day, day of the week, and time of the year \cite{Weron2014}. Due to these facts, the dynamics of electricity prices exhibit behavior unseen in other markets, e.g.~sudden and unexpected price peaks or seasonality of prices at three different levels (daily, weekly, and yearly) \cite{Weron2014}. 

As a result of this unique behavior, electricity markets have become a central point of research in the energy sector and accurate electricity price forecasting has emerged as one of the biggest challenges faced by the different market entities. The usual motivation behind these efforts is a purely economic one: as forecasting accuracy increases, the negative economic effects of price uncertainty are mitigated and the market players make an economic profit. In addition, another important fact to consider is that electricity markets are established to keep the grid stable. In particular, as prices become more volatile, the balance of the grid is compromised, strategic reserves may have to be used, and the risk of a blackout increases. Therefore, by accurately forecasting electricity prices, not only economic profits can be made, but also the system stability is improved.

Due to the above motivations, electricity price forecasting has been continuously developed and improved for the last decades, and as a result, the literature comprises a large variety of distinctive approaches, e.g.~see the literature review \cite{Weron2014}. Nevertheless, to the best of our knowledge, a topic that has been not yet addressed is the influence of neighboring and connected markets, i.e.~market integration, on the forecast accuracy. In particular, as different areas in the world, e.g.~the European Union \cite{Jamasb2005}, are enforcing a larger level of integration across national electricity markets, it is sensible to assume that neighboring markets might play a role in the forecasting efficiency. To address this scientific gap, this paper proposes a modeling framework that is able to improve predictive accuracy by exploiting the relations across electricity markets. In particular, by modeling market integration in two different ways, the proposed framework is shown to obtain statistically significant improvements.

The paper is organized as follows: Section \ref{sec:survey} starts by presenting the literature review, motivation, and contributions. Next, Section \ref{sec:theory} and \ref{sec:data} respectively describe the methods and data that are used in the research. Then, Section \ref{sec:modelframe} defines the proposed modeling framework. Next, Section \ref{sec:feat} derives a novel approach for feature selection and uses it to select the optimal features in the case study. Finally, Section \ref{sec:results} evaluates the proposed modeling framework by means of predictive accuracy, and Section 8 summarizes and concludes the paper.

\section{Literature Survey and Contributions}
\label{sec:survey}
In this section, we present the literature review of three topics that are relevant for the research: electricity price forecasting, market integration, and feature selection. Based on that, we motivate our work and explain our contributions.

\subsection{Electricity Price Forecasting}
{The price forecasting literature is typically divided into five areas: (1) {multi-agent or game theory models} simulating the operation of market agents, (2) {fundamental methods} employing physical and economic factors, (3) {reduced-form models} using statistical properties of electricity trade for risk and derivatives evaluation, (4) {statistical models} comprising time series and econometric models, and (5) {artificial intelligence methods} \cite{Weron2014}. For forecasting day-ahead prices, or in general any other type of electricity spot prices, statistical and artificial intelligence methods have showed to yield the best results \cite{Weron2014}. As a result, they are the main focus of this review.

Typical statistical methods are: {AR} and {ARX models} \cite{Weron2008}, {ARIMA models} \cite{CrespoCuaresma2004,Yang2017}, {dynamic regression} \cite{Nogales2002}, {transfer functions} \cite{Nogales2002}, {double seasonal Holtz-Winter} model \cite{Cruz2011}, {TARX} model \cite{Misiorek2006}, {semi/non-parametric models} \cite{Weron2008}, or {GARCH-based models} \cite{Diongue2009}. In addition, within the same class of methods, different hybrid models have been also applied, e.g.~wavelet-based models \cite{Conejo2005,Tan2010,Yang2017}.

Statistical models are usually linear forecasters, and as such, they are successful in the areas where the frequency of the data is low, e.g.~for weekly patterns. However, for hourly values, the nonlinear behavior of the data might be too complicated to predict  \cite{Amjady2006a}. As a result, motivated by the need for forecasters that are able to predict the nonlinear behavior of hourly prices, several artificial intelligence methods have been proposed. Among these methods, {artificial neural networks} \cite{Szkuta1999,Catalao2007,Xiao2017,Wang2017a}, {support vector regressors} \cite{Fan2007}, {radial basis function networks} \cite{Lin2010}, and {fuzzy networks} \cite{Amjady2006} are among the most commonly used. A recent study \cite{Lago2017} showed that \textit{Deep Neural Networks (DNNs)} can also be a successful alternative.

The results comparing the accuracy of the mentioned models have however produced unclear conclusions \cite{Catalao2007}. In general, the effectiveness of each model seems to depend on the market under study and on the period considered.}

\subsection{Market Integration}
In the last decades, the EU has passed several laws trying to achieve a single and integrated European electricity market \cite{Meeus2008,Jamasb2005}. At the moment, while a single market is far from existing, there is evidence suggesting that the level of integration across the different regional markets has been increasing over time \cite{Bunn2010}. In particular, evidence suggests that in the case of Belgium and France, the spot prices share strong common dynamics \cite{DeMenezes2016}.

While some researchers have evaluated the level of integration of the European markets \cite{Zachmann2008,Bunn2010,DeMenezes2016}, and others have proposed statistical models to evaluate the probability of spike transmissions across EU markets \cite{Lindstroem2012}, the literature regarding market integration to improve forecasting accuracy is rather scarce. To the best of our knowledge, only two other works have taken into account some sort of market integration, namely \cite{Ziel2015} and \cite{Panapakidis2016}.

In particular, \cite{Ziel2015} analyzes the effect of using the day-ahead prices of the \textit{Energy Exchange Austria (EXAA)} on a given day to forecast the prices of other European markets on the same day. Using the fact that for the EXAA market the clearing prices are released before the closure of other European markets, \cite{Ziel2015} models the price dynamics of several European markets and considers the EXAA prices of the same day as part of these models. It is shown that, for certain European markets, using the available prices from the EXAA improves the forecasting accuracy in a statistically significant manner.

Similarly, \cite{Panapakidis2016} considers external price forecasts from other European markets as exogenous inputs of an artificial neural network to predict Italian day-ahead prices. \cite{Panapakidis2016} shows that using the given forecasts the accuracy of their network can be improved from 19.08\% to 18.40\% \textit{mean absolute percentage error (MAPE)}.

\subsection{Feature Selection}
Feature selection is defined as the process to select, for a given model, the subset of important and relevant input variables, i.e.~features.
Typically, three families of methods to perform feature selection exist: \textit{filter}, \textit{wrapper}, and \textit{embedded methods} \cite{Guyon2003}. Filter methods apply some statistical measure to assess the importance of features \cite{Carta2015}. Their main disadvantage is that, as the specific model performance is not evaluated and the relations between features are not considered, they may select redundant information or avoid selecting some important features. Their main advantage is that, as a model does not have to be estimated, they are very fast. By contrast, wrapper methods perform a search across several feature sets, evaluating the performance of a given set by first estimating the prediction model and then using the predictive accuracy of the model as the performance measure of the set \cite{Carta2015}. Their main advantage is that they consider a more realistic evaluation of the performance and interrelations of the features; their drawback is a long computation time. Finally, embedded methods, e.g.~regularization \cite[Chapter 7]{Goodfellow2016}, learn the feature selection at the same time the model is estimated. Their advantage is that, while being less computationally expensive than wrapper methods, they still consider the underlying model. However, as a drawback, they are specific to a learning algorithm, and thus, they cannot always be applied.

Approaches for feature selection in the electricity price forecasting literature vary according to the prediction model used. For time series methods using only prices, e.g.~ARIMA, {autocorrelation plots} \cite{Conejo2005} or the {Akaike information criterion} \cite{Stevenson2001} have been commonly used. In the case of forecasters with explanatory variables, e.g.~neural networks, most researchers have used trial and error or filter methods based on linear analysis techniques: {statistical sensitivity analysis} \cite{Cruz2011,Szkuta1999}, correlation analysis \cite{Rodriguez2004}, or {principal component analysis} \cite{Hong2012}. Since prices display nonlinear dynamics, the mentioned techniques might be limited \cite{Amjady2009}; to address this, nonlinear filter methods such as the {relief algorithm} \cite{Amjady2010} or techniques based on mutual information \cite{Amjady2009,Keles2016,Ghasemi2016} have been proposed. More recently, a hybrid nonlinear filter-wrapper method, which uses mutual information and information content as a first filter step and a real-coded genetic algorithm as a second wrapper step, has been proposed \cite{Abedinia2017}.

\subsection{Motivation and Contributions}
\label{sec:motivation}
While the effects of market integration can dramatically modify the dynamics of electricity prices, there is a lack of a general modeling framework that could model this effect and analyze its impact on the electricity market. To address this gap, in this paper we provide general models to identify these relations and a technique to quantify the importance of market integration. As we will show, understanding these relations is key to improve the accuracy of forecasting models, and thus, to obtain energy systems that are economically more efficient.

The two available papers on market integration in price forecasting, \cite{Ziel2015,Panapakidis2016}, are both limited to the case where the day-ahead prices of neighboring markets are known in advance. While these papers provide a first modeling approach for market integration, the methodologies are very specific and can only be applied in limited situations. In particular, most European electricity markets release their day-ahead prices at the same time, and thus, the prices of neighboring markets cannot be obtained in advance. The only exception to this rule is the EXAA market, which was the object of study of \cite{Ziel2015}. In addition to this limitation, neither \cite{Ziel2015} nor \cite{Panapakidis2016} analyzed the relevance of market integration.

In contrast to \cite{Ziel2015,Panapakidis2016}, we propose a general modeling framework that is able to model and analyze market integration for any given market. In particular, we propose a modeling framework based on DNNs that considers market integration features that are available beforehand in all European markets. Using past prices and publicly available load/generation forecasts in neighboring markets, we propose a first forecaster that models market integration effects on price dynamics.
Next, we propose a second forecaster that further generalizes market integration: besides modeling market integration using input features, the second forecaster also includes the effect in the output space. By simultaneously predicting prices in multiple markets, the proposed forecaster is able to improve the predictive accuracy.

Finally, we also contribute to the field of feature selection algorithms. More specifically, while the feature selection methods for electricity price forecasting proposed in the literature provide good and fast algorithms, they suffer from three main drawbacks: (1) They all \cite{Conejo2005,Cruz2011,Szkuta1999,Rodriguez2004,Hong2012,Amjady2009,Amjady2010,Keles2016,Abedinia2017} perform a filter step where the model performance is not directly considered; therefore, the resulting selected features might be redundant or incomplete. (2) In the case of the algorithms for nonlinear models \cite{Amjady2009,Amjady2010,Keles2016,Abedinia2017}, the inputs have to be transformed to lower-dimensional spaces; as a result, feature information might be lost. (3) While they provide a selection of features, none of these methods computes the relative importance of each feature.

%\begin{itemize}
%	\item They all \cite{Conejo2005,Cruz2011,Szkuta1999,Rodriguez2004,Hong2012,Amjady2009,Amjady2010,Keles2016,Abedinia2017} perform a filter step \textred{where the model performance is not directly considered but a statistical measure is used instead}. As a result, the resulting selected features might be redundant or incomplete.
%	\item In the case of the algorithms for nonlinear models \cite{Amjady2009,Amjady2010,Keles2016,Abedinia2017}, the inputs have to be transformed to lower-dimensional spaces. As a result, feature information might be lost.
%	\item While they provide a selection of features, none of these methods computes the relative importance of each feature.	
%\end{itemize}
To address these issues, we propose a wrapper selection algorithm based on functional ANOVA that directly selects features using nonlinear models and without any feature transformation. While the proposed approach is computationally more expensive than previously proposed methods, it can perform a more accurate feature selection as it avoids transformations, selects the features based on the original model, and computes the individual performance of each feature.

\section{Preliminaries}
\label{sec:theory}
In this section we introduce the theoretical concepts and algorithms that are used and/or modified later on in the paper. 
%\textblue{In particular, Section \ref{sec:dayahead} briefly describes forecasting day-ahead prices. Then, Section \ref{sec:deeplearning} introduces DNNs, the chosen model for predicting prices. Section \ref{sec:hyper} presents hyperparameter optimization and analysis, foundations that are used for the novel feature selection algorithm. Then, Section \ref{sec:metrics} describes the metric used to assess the accuracy and performance of the proposed models. Finally, Section \ref{sec:DM} defines the Diebold-Mariano test, a tool to assess the statistical significance of predictive accuracy.}

\subsection{Day-ahead Forecasting}
\label{sec:dayahead}
The day-ahead electricity market is a type of power exchange widely used in several regions of the world. In its most general format, producers and consumers have to submit bids for the 24 hours of day $d$ before some deadline on day $d-1$ (in most European markets, this deadline occurs at 11:00 am or 12:00 am). Except for some markets, these bids are typically defined per hour, i.e.~every market player has to submit 24 bids.
%with the exception of some markets, e.g.~in UK the bids are per 30 minutes intervals.

After the deadline, the market operator takes into account all the bids and computes the market clearing price for each of the 24 hours. 
Then, consumer/producer bids larger/lower or equal than the market clearing prices are approved, and a contract is established.
%and these consumers (producers) are obliged to satisfy their bids.

A useful forecaster of the day ahead market should thus be able to predict the set of 24 market clearing prices of day $d$ based on the information available before the deadline of day $d-1$.

\subsection{Deep Learning and DNNs}
\label{sec:deeplearning}
 During the last decade, the field of neural networks has gone trough some major innovations that have lead to what nowadays is known as deep learning \cite{Goodfellow2016}. Specifically, the term deep refers to the fact that, thanks to the novel developments of recent years, we can now train different neural network configurations whose depth is not just limited to a single hidden layer (as in the traditional multilayer perceptron), and which have systemically showed better generalization capabilities \cite{Goodfellow2016}.

While there are different DNN architectures, e.g.~convolutional networks or recurrent networks, in this paper we consider a standard DNN, i.e.~a multilayer perceptron with more than a single hidden layer. 

\subsubsection{Representation}
\begin{figure}[htb]
	\begin{center}
		\def\Nin{2}
\def\Nhid{2}
\def\Nhidd{2}
\def\Nout{2}
\setlength{\separ}{2cm}
\begin{tikzpicture}[shorten >=1pt,->,draw=black!50]

\tikzstyle{every pin edge}=[<-,shorten <=1pt]
\tikzstyle{neuron}=[circle,fill=black!25,minimum size=17pt,inner sep=0pt]
\tikzstyle{input neuron}=[neuron, fill=greenPlots!60];
\tikzstyle{output neuron}=[neuron, fill=redPlots!50];
\tikzstyle{hidden neuron}=[neuron, fill=bluePlots!50];
\tikzstyle{hidden neuron 2}=[neuron, fill=orange!50];
\tikzstyle{annot} = [text width=4em, text centered]

% Draw the input layer nodes
\foreach \name / \y in {1,...,\Nin}
% This is the same as writing \foreach \name / \y in {1/1,2/2,3/3,4/4}
\node[input neuron] (I-\name) at (0,-\y) {$x_{\y}$};
\node at (0,-\Nin-1) {$\vdots$};
\node[input neuron] (I-\Nin+1) at (0,-\Nin-2) {$x_{\nin}$};

% Draw the hidden layer nodes
\foreach \name / \y in {1,...,\Nhid}
\path[yshift=0cm]
node[hidden neuron] (H-\name) at (\separ,-\y cm) {$\hid_{1\y}$};
\node at (\separ,-\Nhid-1) {$\vdots$};
\node[hidden neuron] (H-\Nhid+1) at (\separ,-\Nhid-2) {$\hid_{1\nhidn_1}$};

% Draw the hidden layer nodes
\foreach \name / \y in {1,...,\Nhidd}
\path[yshift=0cm]
node[hidden neuron 2] (H2-\name) at (2\separ,-\y cm) {$\hid_{2\y}$};
\node at (2\separ,-\Nhidd-1) {$\vdots$};
\node[hidden neuron 2] (H2-\Nhidd+1) at (2\separ,-\Nhidd-2) {$\hid_{2 \nhidn_2}$};

\foreach \name / \y in {1,...,\Nout}
\path[yshift=0cm]
node[output neuron] (O-\name) at (3\separ,-\y cm) {$y_{\y}$};
\node at (3\separ,-\Nout-1) {$\vdots$};
\node[output neuron] (O-\Nout+1) at (3\separ,-\Nout-2) {$y_{{\nout}}$};

% Draw the output layer node

% Connect every node in the input layer with every node in the
% hidden layer.

%Input to hidden
\foreach \source in {1,...,\Nin}
\foreach \dest in {1,...,\Nhid}
\path (I-\source) edge (H-\dest);

\foreach \dest in {1,...,\Nhid}
\path (I-\Nin+1) edge (H-\dest);
\foreach \source in {1,...,\Nin}
\path (I-\source) edge (H-\Nhid+1);
\path (I-\Nin+1) edge (H-\Nhid+1);

%Hidden to hidden
\foreach \source in {1,...,\Nhid}
\foreach \dest in {1,...,\Nhidd}
\path (H-\source) edge (H2-\dest);

\foreach \dest in {1,...,\Nhidd}
\path (H-\Nhid+1) edge (H2-\dest);
\foreach \source in {1,...,\Nhid}
\path (H-\source) edge (H2-\Nhidd+1);
\path (H-\Nhid+1) edge (H2-\Nhidd+1);

%Hidden to output
\foreach \source in {1,...,\Nhidd}
\foreach \dest in {1,...,\Nout}
\path (H2-\source) edge (O-\dest);

\foreach \dest in {1,...,\Nout}
\path (H2-\Nhidd+1) edge (O-\dest);
\foreach \source in {1,...,\Nhid}
\path (H2-\source) edge (O-\Nout+1);
\path (H2-\Nhidd+1) edge (O-\Nout+1);

% Annotate the layers
\node[annot,above of=H-1, node distance=0.75cm] (hl) {Hidden layer};
\node[annot,above of=H2-1, node distance=0.75cm] (hl) {Hidden layer};
\node[annot,above of=I-1, node distance=0.75cm] {Input layer};
\node[annot,above of=O-1, node distance=0.75cm] {Output layer};

\node at (3\separ,-\Nout-1) {$\vdots$};
\end{tikzpicture}
		\caption{Example of a DNN.}
		\label{fig:hiddnetexa}
	\end{center}
\end{figure}
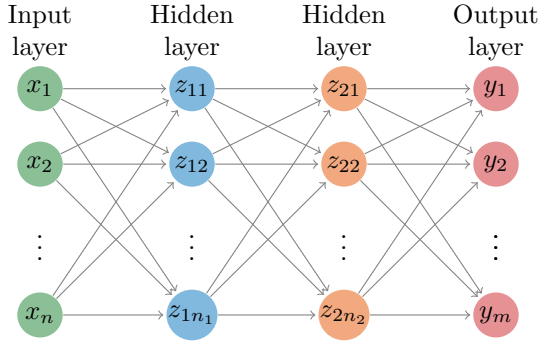

Defining by $\mathbf{X}=[x_1,\ldots,x_\nin]^\top\in\R^\nin$ the input of the network, by $\mathbf{Y}=[y_{{1}},y_{{2}},\ldots,y_{{\nout}}]^\top\in\R^{\nout}$ the output of the network, by $\nhidn_\ixhid$ the number of neurons of the $\ixhid^\mr{th}$ hidden layer, and by $\mathbf{\hid}_\ixhid=[\hid_{\ixhid1},\ldots,\hid_{\ixhid \nhidn_\ixhid}]^\top$ the state vector in the ${\ixhid}^\mr{th}$ hidden layer, a general DNN with two hidden layers can be represented as in Figure \ref{fig:hiddnetexa}.

In this representation, the parameters of the model are represented by the set of weights $\vc{W}$ that establish the mapping connections between the different neurons of the network \cite{Goodfellow2016}.

\subsubsection{Training}

The process of estimating the model weights $\vc{W}$ is usually called training. In particular, given a training set $\mathcal{S_T}=\bigl\{(\X_\ixts,\vc{Y}_\ixts)\bigr\}_{\ixts=1}^\nts$ with $N$ data points, the network training is done by solving a general optimization problem with the following structure:
\begin{mini}
	{\vc{W}}{\sum_{\ixts=1}^{\nts}g_\ixts\Bigl(\vc{Y}_\ixts, F(\vc{X}_\ixts,\vc{W})\Bigr),}
	{\label{eq:prob1}}{}
\end{mini}

\noindent where $F:\R^\nin\rightarrow\R^{\nout}$ is the neural network map, and $g_\ixts$ is the problem-specific cost function, e.g.~the Euclidean norm or the average cross-entropy. Traditional methods to solve \eqref{eq:prob1} include \textit{gradient descent} or the \textit{Levenberg–Marquardt} algorithm \cite{Weron2014}. However, while these methods work well for small sized-networks, they display computational and scalability issues for DNNs. In particular, better alternatives for DNNs are the \textit{stochastic gradient descent} and all its variants \cite{Ruder2016}.

{It is important to note that \eqref{eq:prob1} is an approximation of the real problem we wish to minimize. Particularly, in an ideal situation, we would minimize the cost function w.r.t.~to the underlying data distribution; however, as the distribution is unknown, the problem has to be approximated by minimizing the cost function over the finite training set. This is especially relevant for neural networks, where a model could be overfitted and have a good performance in the training set, but perform badly in the test set, i.e.~a set with a different data distribution. To avoid this situation, the network is usually trained in combination with regularization techniques, e.g.~early stopping, and using out-of-sample data to evaluate the performance \cite{Goodfellow2016}.}

\subsubsection{Network Hyperparameters}
In addition to the weights, the network has several parameters that need to be selected before the training process. Typical parameters include the number of neurons of the hidden layers, the number of hidden layers, the type of activation functions, or the learning rate of the stochastic gradient descent method. To distinguish them from the main parameters, i.e.~the network weights, they are referred to as the network hyperparameters.

\subsection{Hyperparameter Selection}
\label{sec:hyper}
In order to perform the selection of model hyperparameters, papers in the field of electricity price forecasting have traditionally defined a number of configurations and chosen the one with the best performance \cite{Cruz2011,Catalao2007,Amjady2009,Panapakidis2016,Rodriguez2004}. Another approach, yet less usual, has been the use of evolutionary optimization algorithms in order to select the best network configuration \cite{Shafie2011}. However, while these approaches might work under some conditions, they have some flaws. In particular, while the first method implements fast decision-making, it does not provide an optimal selection of hyperparameters. Similarly, while the second method optimizes the selection, it evaluates a very large number of points in the hyperparameter space. As a result, if the function to be evaluated is costly, e.g.~when training a DNN, the second method requires a large computation time.

An alternative to tackle these issues is \textit{Bayesian optimization} \cite{Jones1998}, a family of algorithms for optimizing black-box functions that require a lower number of function evaluations than evolutionary optimization techniques. In particular, their working principle is to sequentially evaluate new samples in the function space, drawing new samples by using the information obtained in the previously explored samples as a prior belief. Based on that, they reduce the number of evaluated sample points and lead to a more efficient optimization.

\subsubsection{Hyperparameter Optimization}
We consider a Bayesian optimization algorithm that has been widely used in the machine learning community. In particular, we use the \textit{Tree-Structured Parzen Estimator (TPE)} \cite{Bergstra2011}, an optimization algorithm within the family of \textit{sequential model-based optimization} methods \cite{Hutter2011}. The basic principle of a sequential model-based optimization algorithm is to optimize a black-box function, e.g.~the performance of a neural network as a function of the hyperparameters, by iteratively estimating an approximation of the function and exploring the function space using the local minima of the approximation. At any given iteration $i$, the algorithm evaluates the black-box function at a new point $\bm{\theta}_i$. Next, it estimates an approximation $\mathcal{M}$ of the black-box function by fitting the previously sampled points to the obtained function evaluations. Then, it selects the next sample point $\bm{\theta}_{i+1}$ by numerically optimizing $\mathcal{M}$ and starts the next iteration. Finally, after a maximum number of iterations $\nruns$ have been performed, the algorithm selects the best configuration. 	
%In the case of a neural network and its hyperparameters, the algorithm evaluates the performance of a hyperparameter instantiation $\bm{\theta}_i$ by training the neural network and computing its predictive accuracy $p_i$. Then, it estimates a model $\mathcal{M}$ that fits all the past hyperparameter instantiations $\bm{\theta}_i$ to the respective performances $p_i$, and selects the next instantiation $\bm{\theta}_{i+1}$ by numerically optimizing $\mathcal{M}$.
Algorithm \ref{al:smbo} represents an example of a sequential model-based optimization algorithm for hyperparameter selection.
\begin{algorithm}
	\caption{Hyperparameter Optimization}
	\label{al:smbo}
	\begin{algorithmic}[1]
		\Procedure{SMBO}{$\nruns,\bm{\theta}_0$}
		\State $\bm{\theta}_i \gets \bm{\theta}_0$
		\State $\mathcal{H} \gets \emptyset $
		\For{$i=1,\ldots,\nruns$}
		\State $ p_i \gets$ TrainNetwork($\bm{\theta}_i$) %\Comment{\#$p$ performance}
		\State $\mathcal{H} \gets \mathcal{H} \cup \bigl\{(p_i,\bm{\theta}_i)\bigr\}$
		%				\State $\mathcal{H} \gets \mathcal{H}+\!\!\!+(p_i,\theta_i)$
		\If{$i < T$}
		\State $\mathcal{M}_i(\bm{\theta}) \gets \mr{EstimateModel}(\mathcal{H})$
		\State $ \bm{\theta}_i \gets \mr{argmax}_{\bm{\theta}} ~\mathcal{M}_i(\bm{\theta})$
		\EndIf
		\EndFor
		\State $\bm{\theta}^* \gets \mr{BestHyperparameters}(\mathcal{H})$
		\State \Return $\bm{\theta}^*$
		\EndProcedure
	\end{algorithmic}
\end{algorithm}

\subsubsection{Hyperparameter Analysis}
\label{sec:fanova}
An optional step after hyperparameter optimization is to perform an analysis of the hyperparameter importance. In particular, while the optimal hyperparameter configuration has been already obtained, it is unknown how much each hyperparameter contributes to the overall performance. Investigating this is specially relevant in order to avoid unnecessary model complexities; e.g.~while the optimal number of neurons might be large, reducing the number of neurons might barely affect the performance. 
%\textblue{In that case, it might be better to use a smaller network and improve the algorithm computational speed at the cost of a slightly lower performance.}

\paragraph{Functional ANOVA}
An approach for carrying on such an analysis is proposed in \cite{Hutter2014}, where a novel method based on random forests and functional ANOVA is introduced. In particular, \cite{Hutter2014} considers the generic case of having $\nhyper$ hyperparameters with domains $\Theta_1, \ldots, \Theta_\nhyper$, and defines the following concepts:

\begin{itemize}
	\item Hyperparameter set $\hyperset=\{1,\ldots,\nhyper\}$. % i.e~the hyperparameters are defined by integers.
	\item Hyperparameter space $\vc{\Theta}:\Theta_1 \times \ldots \times \Theta_\nhyper$.
	\item Hyperparameter instantiation $\bm{\theta}=[\theta_1,\ldots,$ $\theta_\nhyper]^\top$.
	\item Hyperparameter subset $U=\{u_{1},\ldots,u_{\nhypersub}\} \subseteq \hyperset$ and associated
	partial hyperparameter instantiation $\bm{\theta}_U=[\theta_{u_1},\ldots,\theta_{u_\nhypersub}]^\top$.
\end{itemize}

Then, given a set $\mathcal{H}=\bigl\{(\boldsymbol{\theta}_\ixts,p_\ixts  )\bigr\}_{\ixts=1}^\nruns$ of hyperparameter realizations, the proposed method fits a random forest model $\mathcal{M}_\mr{RF}(\boldsymbol{\theta})$ to build a predictor of the performance $p$ as a function of the hyperparameter vector $\boldsymbol{\theta}$.  

Then, using $\mathcal{M}_\mr{RF}$, the method defines a \textit{marginal performance predictor} $\hat{a}(\boldsymbol{\theta}_U)$ as a forecaster of the performance of any partial hyperparameter instantiation $\boldsymbol{\theta}_U$. In particular, given a subset $U\subseteq \hyperset$, $\hat{a}(\boldsymbol{\theta}_U)$ provides an estimation of the average performance across the hyperparameter space $\hyperset\setminus U$  when the hyperparameters of $U$ are fixed at $\boldsymbol{\theta}_U$. 

Finally, using the marginal performance predictor $\hat{a}(\boldsymbol{\theta}_U)$, the algorithm carries out a functional ANOVA analysis to estimate the importance of each hyperparameter. Particularly, defining the total variance across the performance by $\mathbb{V}$, the algorithm partitions $\mathbb{V}$ as a sum of individual variance contributions of subsets $U\subseteq \hyperset$ to $\mathbb{V}$  :
\begin{equation}
\label{eq:variance}
\mathbb{V} =\sum_{U\subseteq \hyperset} \mathbb{V}_U,
\end{equation}
\noindent where $\mathbb{V}_U$ is the contribution of subset $U$ to the total variance. Then, the importance $\mathbb{F}_U$ of each subset $U$ is computed based on the subset contribution to the total performance variance:
\begin{equation}
\label{eq:importance}
\mathbb{F}_U = \frac{\mathbb{V}_{U}}{\mathbb{V}}.
\end{equation}
\noindent For the particular case of the hyperparameter importance, the algorithm just evaluates $\mathbb{F}_U$ for each subset $U=\{ \ixset \}$ composed of a single hyperparameter. As in \cite{Hutter2014}, we refer to the variance contributions $\mathbb{F}_U$ of single hyperparameters as \textit{main effects} and to the rest as \textit{interaction effects}.

It is important to note that, in addition to the importance $\mathbb{F}_U$, the algorithm also provides, for each partial hyperparameter instantiation $\boldsymbol{\theta}_U$, the prediction of the marginal performance $\hat{a}(\boldsymbol{\theta}_U)$ and an estimation of its standard deviation $\sigma_{\boldsymbol{\theta}_U}$.

\subsection{Performance Metrics}
\label{sec:metrics}
In order to evaluate the accuracy of the proposed models, we need a performance metric. In this paper, as motivated below, we use the \textit{symmetric mean absolute percentage error (sMAPE)} \cite{Makridakis1993}. Given a vector $\vc{Y}=[y_1,\ldots,y_\nts]^\top$ of real outputs and a vector $\vc{\hat{Y}}=[\hat{y}_1,\ldots,\hat{y}_\nts]^\top$ of predicted outputs, the sMAPE metric can be computed as:
\begin{equation}
\mathrm{sMAPE} = \frac{100}{\nts}\sum_{\ixts=1}^{\nts} \frac{|y_\ixts-\hat{y_\ixts}|}{(|y_\ixts|+|\hat{y}_\ixts|)/2}.
\end{equation}
The reason for selecting the sMAPE instead of the more traditional MAPE is the fact that the MAPE is affected by different issues \cite{Makridakis1993}. Particularly, for our application, the MAPE becomes sensitive to values close to zero. When an output $y_i$ gets close to zero, the corresponding MAPE contribution becomes very large and it dominates the final value.

\subsection{Diebold-Mariano (DM) Test}
\label{sec:DM}
The sMAPE metric defined above only provides an assessment of which model has, for the data use, a better accuracy. While the accuracy of a model can be higher, the difference in performance might be not significant enough to establish that the model is really better. To assess the statistical significance in the difference of predictive accuracy performance, a commonly used tool is the Diebold-Mariano test \cite{Diebold1995}. 

Given a time series vector $\vc{Y}=[y_1,\ldots,y_\nts]^\top$ to be forecasted, two prediction models $M_1$ and $M_2$, and the associated forecasting errors $\bm{\varepsilon}^{M_1}=[\varepsilon_1^{M_1},\ldots,\varepsilon_\nts^{M_1}]^\top$ and $\bm{\varepsilon}^{M_2}=[\varepsilon_1^{M_2},\ldots,\varepsilon_\nts^{M_2}]^\top$, the DM test evaluates whether there is a significant difference in performance accuracy based on an error loss function $L(\varepsilon_\ixts^{M_i})$. In particular, the DM test builds a loss differential function as:
\begin{equation}
\label{eq:lossdif}
d^{M_1,M_2}_k = L(\varepsilon_\ixts^{M_1}) - L(\varepsilon_\ixts^{M_2}),
\end{equation}

\noindent and then, it tests the null hypothesis $H_0$ of both models having equal accuracy, i.e.~equal expected loss, against the alternative hypothesis $H_1$ of the models having different accuracy, i.e.:
\begin{equation}
\label{eq:dsdm}
\begin{tabular}{c}
Two-sided\\DM test
\end{tabular}
\begin{cases}
H_0:~\mathbb{E}(d^{M_1,M_2}_k)=0,\\
H_1:~\mathbb{E}(d^{M_1,M_2}_k)\neq0,
\end{cases}
\end{equation}

\noindent with $\mathbb{E}$ representing the expected value. Similar to the standard two-sided test, a one-sided DM test can be built by testing the null hypothesis that the accuracy of $M_1$ is equal or worse than the accuracy of $M_2$ versus the alternative hypothesis of the accuracy of $M_1$ being better:
\begin{equation}
\label{eq:osdm}
\begin{tabular}{c}
One-sided\\DM test
\end{tabular}
\begin{cases}
H_0:~\mathbb{E}(d^{M_1,M_2}_k)\geq0,\\
H_1:~\mathbb{E}(d^{M_1,M_2}_k)<0.
\end{cases}
\end{equation}

While the loss function $L$ can be freely chosen, it has to ensure that the resulting loss differential is covariance stationary. A loss function that is typically used is:
\begin{equation}
\label{eq:lf}
L(\varepsilon_\ixts^{M_i}) = |\varepsilon_\ixts^{M_i}|^p,
\end{equation}
\noindent where usually $p\in\{1,2\}$.

\section{Data}
\label{sec:data}
In this section, the data used for the research is introduced.

\subsection{Data Selection and Motivation}
In general, when looking at the day-ahead forecasting literature, many inputs have been proposed as meaningful explanatory variables, e.g.~temperature, gas and coal prices, grid load, available generation, or weather \cite{Weron2014}.

To make our selection, we try to make sure that the selected data is not only related to the price dynamics, but also fulfills some minimum requirements. More specifically, we only choose data that is freely available for most European markets so that the proposed models can easily be exported to other EU markets. Moreover, we ensure that the data represents market integration, i.e.~that comes from two connected markets. In particular, we select the period from 01/01/2010 to 31/11/2016 as the time range of study, and we consider the following data:

\begin{enumerate}
	\item Day-ahead prices from the EPEX-Belgium and EPEX-France power exchanges. They are respectively denoted as $p_\mr{B}$ and $p_\mr{F}$.
	\item Day-ahead forecasts of the grid load and generation capacity in Belgium and France. Like in other European markets, these forecasts are available before the bid deadline on the website of the \textit{transmission system operators (TSOs)}: ELIA for Belgium and RTE for France. They are respectively denoted as $l_\mr{B}$ and $g_\mr{B}$ for Belgium, and as $l_\mr{F}$ and $g_\mr{F}$ for France.
	\item Calendar of public holidays $H_F$ and $H_B$ in France and Belgium in the defined time range.
\end{enumerate}

While it could be argued that different weather data could also be easily accessible and important for the forecasting, for our research, we have decided to disregard them for two main reasons:
\begin{enumerate}
	\item Weather factors are already indirectly taken into account in the grid load and generation forecasts provided by the TSO. In particular, the generation forecast has to consider weather information regarding wind speed and solar radiation. Likewise, load forecasts also need to consider temperature and other weather variables to obtain the electricity consumption.
	%	\item In addition, while it is true that obtaining weather forecasts is relatively easy, acquiring historical forecast data turns out to be more difficult. In particular, free services limit the number of request per day, which in turns makes the process of obtaining years of historical data into a very lengthy procedure.
	\item Weather data are local phenomena, and as such, they can greatly vary from one part of a country to another. As a result, unlike the grid load or generation data, it is not possible to select a single value of the temperature or any other weather data for a given time interval.
\end{enumerate}
%Based on the above arguments, we decide that it is more sensible to consider only grid load and available capacity as, first, both are freely accessible across the different European markets, second, they also consider some weather information data, and third, they are represented by a unique value per time interval.

\subsection{Data Processing}
It is important to note that the data used is mostly unprocessed. In particular, as we intend to forecast and detect spikes, price outliers are not eliminated. The only data transformation is a price interpolation and elimination every year corresponding respectively to the missing and extra values due to the daylight saving. In addition, while all the metrics and tests are computed using the real prices, the training of the neural networks is done with data normalized to the interval $[-1,1]$. This last step is necessary because the input features have very different ranges; therefore, if the data is not normalized, the training time increases and the final result is a network that displays, in general, worse performance \cite{LeCun1998}.

\subsection{Data Division}
To perform the different experiments, we divide the data into three sets:

\begin{enumerate}
	\item Training set (01/01/2010 to 31/11/2014): These data are used for training and estimating the different models.
	\item Validation set (01/11/2014 to 31/11/2015): A year of data is used to conduct early-stopping to ensure that the model does not overfit and to select optimal hyperparameters and features.
	\item Test set (01/11/2015 to 31/11/2016): A year of data, which is not used at any step during the model estimation process, is employed as the out-of-sample dataset to compare and evaluate the models.
\end{enumerate}

\subsection{Data Access}
For the sake of reproducibility, we have only used publicly available data. In particular, the load and generation day-ahead forecasts are available on the webpages of RTE \cite{RTE} and Elia \cite{Elia}, the respective TSOs in France and Belgium. In the case of the prices, they can be obtained from the ENTSO-E transparency platform \cite{entsoe}.

\section{Modeling Framework}
\label{sec:modelframe}
In this section, two different models are proposed to include market integration in day-ahead forecasting. The two models are similar to each other as both of them try to forecast the full set of day-ahead prices. However, they differ from each other in the number and type of prices that they predict; in particular, while the first model predicts the day-ahead prices of a single market, the second model combines a dual market prediction into a single model. 

\subsection{Single-Market Day-Ahead Forecaster}
\label{sec:dayaheadfor}
The basic model for predicting day-ahead prices uses a DNN in order to forecast the set of 24 day-ahead prices.
\subsubsection{Conceptual Idea}
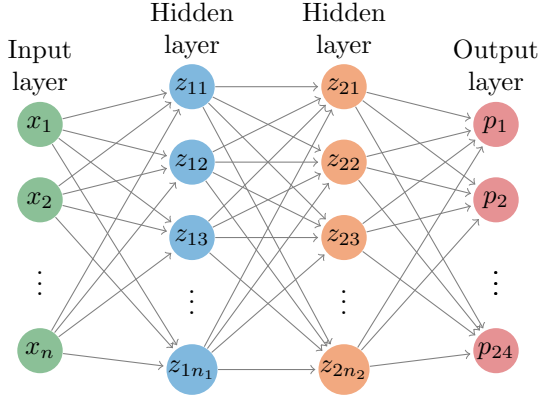
\begin{figure}[htb]
	\begin{center}
\def\Nin{2}
\def\Nhid{3}
\def\Nhidd{3}
\def\Nout{2}
%\newlength{\separ}
\setlength{\separ}{2cm}
\begin{tikzpicture}[shorten >=1pt,->,draw=black!50]

\tikzstyle{every pin edge}=[<-,shorten <=1pt]
\tikzstyle{neuron}=[circle,fill=black!25,minimum size=17pt,inner sep=0pt]
\tikzstyle{input neuron}=[neuron, fill=greenPlots!60];
\tikzstyle{output neuron}=[neuron, fill=redPlots!50];
\tikzstyle{hidden neuron}=[neuron, fill=bluePlots!50];
\tikzstyle{hidden neuron 2}=[neuron, fill=orange!50];
\tikzstyle{annot} = [text width=4em, text centered]

% Draw the input layer nodes
\foreach \name / \y in {1,...,\Nin}
% This is the same as writing \foreach \name / \y in {1/1,2/2,3/3,4/4}
\node[input neuron] (I-\name) at (0,-\y) {$x_{\y}$};
\node at (0,-\Nin-1) {$\vdots$};
\node[input neuron] (I-\Nin+1) at (0,-\Nin-2) {$x_{\nin}$};

% Draw the hidden layer nodes
\foreach \name / \y in {1,...,\Nhid}
\path[yshift=0.5cm]
node[hidden neuron] (H-\name) at (\separ,-\y cm) {$\hid_{1\y}$};
\node at (\separ,-\Nhid-0.25) {$\vdots$};
\node[hidden neuron] (H-\Nhid+1) at (\separ,-\Nhid-1.25) {$\hid_{1\nhidn_1}$};

% Draw the hidden layer nodes
\foreach \name / \y in {1,...,\Nhidd}
\path[yshift=0.5cm]
node[hidden neuron 2] (H2-\name) at (2\separ,-\y cm) {$\hid_{2\y}$};
\node at (2\separ,-\Nhidd-0.25) {$\vdots$};
\node[hidden neuron 2] (H2-\Nhidd+1) at (2\separ,-\Nhidd-1.25) {$\hid_{2\nhidn_2}$};

\foreach \name / \y in {1,...,\Nout}
\path[yshift=0cm]
node[output neuron] (O-\name) at (3\separ,-\y cm) {$p_{\y}$};
\node at (3\separ,-\Nout-1) {$\vdots$};
\node[output neuron] (O-\Nout+1) at (3\separ,-\Nout-2) {$p_{{24}}$};

% Draw the output layer node

% Connect every node in the input layer with every node in the
% hidden layer.

%Input to hidden
\foreach \source in {1,...,\Nin}
\foreach \dest in {1,...,\Nhid}
\path (I-\source) edge (H-\dest);

\foreach \dest in {1,...,\Nhid}
\path (I-\Nin+1) edge (H-\dest);
\foreach \source in {1,...,\Nin}
\path (I-\source) edge (H-\Nhid+1);
\path (I-\Nin+1) edge (H-\Nhid+1);

%Hidden to hidden
\foreach \source in {1,...,\Nhid}
\foreach \dest in {1,...,\Nhidd}
\path (H-\source) edge (H2-\dest);

\foreach \dest in {1,...,\Nhidd}
\path (H-\Nhid+1) edge (H2-\dest);
\foreach \source in {1,...,\Nhid}
\path (H-\source) edge (H2-\Nhidd+1);
\path (H-\Nhid+1) edge (H2-\Nhidd+1);

%Hidden to output
\foreach \source in {1,...,\Nhidd}
\foreach \dest in {1,...,\Nout}
\path (H2-\source) edge (O-\dest);

\foreach \dest in {1,...,\Nout}
\path (H2-\Nhidd+1) edge (O-\dest);
\foreach \source in {1,...,\Nhid}
\path (H2-\source) edge (O-\Nout+1);
\path (H2-\Nhidd+1) edge (O-\Nout+1);

% Annotate the layers
\node[annot,above of=H-1, node distance=0.75cm] (hl) {Hidden layer};
\node[annot,above of=H2-1, node distance=0.75cm] (hl) {Hidden layer};
\node[annot,above of=I-1, node distance=0.75cm] {Input layer};
\node[annot,above of=O-1, node distance=0.75cm] {Output layer};

\node at (3\separ,-\Nout-1) {$\vdots$};
\end{tikzpicture}
\caption{DNN to forecast day-ahead prices.}
\label{fig:hiddnet}
\end{center}
\end{figure}

Based on the results of \cite{Lago2017}, we select a DNN with two hidden layers as forecasting model. 
Defining the input of the model as the relevant data $\mathbf{X}=[x_1,\ldots,x_\nin]^\top\in\R^\nin$ available at day $d-1$ in the local and neighboring markets, and letting $\nhidn_1$ and $\nhidn_2$ be the number of neurons of the first and the second hidden layer respectively, and
$\mathbf{p}=[p_{{1}},p_{{2}},\ldots,p_{{24}}]^\top\in\R^{24}$ the set of 24 day-ahead prices to be forecasted, the proposed model can be represented as in Figure \ref{fig:hiddnet}. 
%The equations of the model are given by \eqref{eq:deepnn}, where in our case the number of hidden layers is $\nhid=2$ and the output dimension is $\nout=24$.

\subsubsection{Model Parameters}
The parameters of the DNN are represented by the set of weights that establish the mapping connections between the different neurons of the network:

\begin{itemize}
	\item $\mathbf{W}_{\mr{i},\ixneuron}$: the vector of weights between the input $\X$ and the neuron $\ixneuron$ of the first hidden layer.
	\item $\mathbf{W}_{{\mr{h}, \ixneuron}}$: the vector of weights between the first hidden layer and the neuron $\ixneuron$ of the second hidden layer.
	\item $\mathbf{W}_{\mr{o},\ixneuron}$: the vector of weights between the second hidden layer and the output price vector $\mathbf{p}$.
	\item $\mathbf{b}_\ixhid =[b_{\ixhid1},\ldots,b_{\ixhid{\nhidn_\ixhid}}]^\top$: the vector of bias weights in the ${\ixhid }^\mr{th}$ hidden layer, with $k=1,2$.
	\item $\mathbf{b}_\mr{o}=[b_{\mr{o},1}\ldots,b_{\mr{o},24}]^\top$: the vector of bias weights in the output layer.
\end{itemize}

\subsubsection{Model Equations}
Using the above definitions, the equations of the DNN can be defined as:
\begin{subequations}
	\label{eq:deepnn}
	\begin{alignat}{3}
	\!\!\!\!\hid_{1\ixneuron} &= f_{1\ixneuron}\Bigl(\mathbf{W}_{\mr{i},\ixneuron}^\top \cdot \mathbf{X}+b_{1\ixneuron}\Bigr),\quad &&\mr{for~}\ixneuron=1,\ldots \nhidn_1,\\
	%	\hid_{2\ixneuron} &= f_{2\ixneuron}\Bigl(\mathbf{W}_{{2\ixneuron}}^\top \cdot  \mathbf{\hid}_1+b_{2\ixneuron}\Bigr),\quad &&\forall \ixneuron=1,\ldots \nhidn_2,\\
	%& ~~\vdots \nonumber\\
	\!\!\!\!\hid_{2 \ixneuron} &= f_{2 \ixneuron}\Bigl(\mathbf{W}_{{\mr{h},\ixneuron}}^\top \cdot  \mathbf{\hid}_{1}+b_{2\ixneuron}\Bigr), ~&&\mr{for~} \ixneuron=1,\ldots \nhidn_2,\\
	%	& ~~\vdots \nonumber\\
	\!\!\!\! p_{\ixneuron} &= f_{\mr{o},\ixneuron}\Bigl(\mathbf{W}_{\mr{o},\ixneuron}^\top \cdot \mathbf{\hid}_2+b_{\mr{o},\ixneuron}\Bigr),\quad &&\mr{for~} \ixneuron=1,\ldots 24, \label{eq:genoutDNN}
	\end{alignat}
\end{subequations}

\noindent where $f_{1 \ixneuron}$ and $f_{2 \ixneuron}$ respectively represent the activation function of neuron $\ixneuron$ in the first and second hidden layer, and where $f_{\mr{o},\ixneuron}$ is the activation function of neuron $\ixneuron$ in the output layer. 

\subsubsection{Network Structure}
The rectified linear unit \cite{Nair2010} is selected as the activation function of the two hidden layers. However, as the prices are real numbers, no activation function is used for the output layer. 

To select the dimension $\nin$ of the network input and the dimensions $\nhidn_1$ and $\nhidn_2$ of the hidden layers, a feature selection and hyperparameter optimization are performed.

%To select these activation functions, the framework considers  activation functions are the sigmoid function, the hyperbolic tangent function, or the rectified linear unit \cite{Goodfellow2016}.

\subsubsection{Training}
\label{sec:singledetails}
The DNN is trained  by minimizing the mean absolute error. In particular, given the training set $\mathcal{S_T}=\bigl\{(\X_\ixts,\vc{p}_\ixts)\bigr\}_{\ixts=1}^\nts$, the optimization problem that is solved to train the neural network is:
\begin{mini}
{\vc{W}}{\sum_{\ixts=1}^{\nts}\|\vc{p}_\ixts - F(\vc{X}_\ixts,\vc{W})  \|_1,}
{}{}
\end{mini}

\noindent where $F:\R^\nin\rightarrow\R^{24}$ is the neural network map. The selection of the mean absolute error instead of the more traditional root mean square error is done for a simple reason: as the electricity prices have very large spikes, the Euclidean norm would put too much importance on the spiky prices.

{The optimization problem is initialized via single-start with the Glorot initialization \cite{Glorot2010} and solved using Adam \cite{Kingma2014}, a version of the stochastic gradient descent method that computes adaptive learning rates for each model parameter. Adam is selected for a clear reason: as the learning rate is automatically computed, the time needed to tune the learning rate is smaller in comparison with other optimization methods. Together with Adam, the forecaster also considers early stopping \cite{Yao2007} to avoid overfitting.}

\subsection{Dual Market Day-Ahead Forecaster}
\label{sec:dayaheadfordual}
A possible variant of the single-market model is a forecaster that predicts the prices of two markets in a single model. While this might seem counter-intuitive at first, i.e.~adding extra outputs to the model could compromise its ability to forecast the set of 24 prices that we are really interested in, this approach can, in fact, lead to neural networks that are able to generalize better.

\subsubsection{Conceptual Idea}
The general idea behind forecasting two markets together is that, as we expect prices in both markets to be interrelated and to have similar dynamics, by forecasting both time series in a single model we expect the neural network to learn more accurate relations. In particular, it has been empirically shown that DNNs can learn features that can, to some extent, generalize across tasks \cite{Yosinski2014}. Similarly, it has also been shown that, by forcing DNNs to learn auxiliary related tasks, the performance and learning speed can be improved \cite{Jaderberg2016,Li2016}.

There are some possible hypotheses that can explain why training with multiple outputs can help to improve the performance:

\begin{enumerate}
\item The simplest explanation is the amount of data: as more data is available, the neural network can learn more relevant features. Moreover, as the tasks are related, the neural network has more data to learn features that are common to all tasks.
\item A second reason is regularization: By solving different tasks, the network is forced to learn features useful for all tasks and to not overfit to the data of a single task.
\end{enumerate}

\subsubsection{Model Implementation}
\begin{figure}[htb]
	\begin{center}	
	\def\Nin{4}
\def\Nhid{5}
\def\Nhidd{5}
\def\Nout{1}
\setlength{\separ}{2cm}
\begin{tikzpicture}[shorten >=1pt,->,draw=black!50]

\tikzstyle{every pin edge}=[<-,shorten <=1pt]
\tikzstyle{neuron}=[circle,fill=black!25,minimum size=17pt,inner sep=0pt]
\tikzstyle{input neuron}=[neuron, fill=greenPlots!60];
\tikzstyle{output neuron}=[neuron, fill=redPlots!50];
\tikzstyle{hidden neuron}=[neuron, fill=bluePlots!50];
\tikzstyle{hidden neuron 2}=[neuron, fill=orange!50];
\tikzstyle{annot} = [text width=4em, text centered]

% Draw the input layer nodes
\foreach \name / \y in {1,...,\Nin}
% This is the same as writing \foreach \name / \y in {1/1,2/2,3/3,4/4}
\node[input neuron] (I-\name) at (0,-\y) {$x_{\y}$};
\node at (0,-\Nin-1) {$\vdots$};
\node[input neuron] (I-\Nin+1) at (0,-\Nin-2) {$x_{\nin}$};

% Draw the hidden layer nodes
\foreach \name / \y in {1,...,\Nhid}
\path[yshift=0.5cm]
node[hidden neuron] (H-\name) at (\separ,-\y cm) {$\hid_{1\y}$};
\node at (\separ,-\Nhid-0.25) {$\vdots$};
\node[hidden neuron] (H-\Nhid+1) at (\separ,-\Nhid-1.25) {$\hid_{1\nhidn_1}$};

% Draw the hidden layer nodes
\foreach \name / \y in {1,...,\Nhidd}
\path[yshift=0.5cm]
node[hidden neuron 2] (H2-\name) at (2\separ,-\y cm) {$\hid_{2\y}$};
\node at (2\separ,-\Nhidd-0.25) {$\vdots$};
\node[hidden neuron 2] (H2-\Nhidd+1) at (2\separ,-\Nhidd-1.25) {$\hid_{2\nhidn_2}$};

\foreach \name / \y in {1,...,\Nout}
\path[yshift=0cm]
node[output neuron] (O-\name) at (3\separ,-\y cm) {$p_{\mr{B}_\y}$};
\node at (3\separ,-\Nout-1) {$\vdots$};
\node[output neuron] (O-\Nout+1) at (3\separ,-\Nout-2) {$p_{\mr{B}_{24}}$};
\node[output neuron] (O-\Nout+2) at (3\separ,-\Nout-3) {$p_{\mr{F}_1}$};
\node at (3\separ,-\Nout-4) {$\vdots$};
\node[output neuron] (O-\Nout+3) at (3\separ,-\Nout-5) {$p_{\mr{F}_{24}}$};
% Draw the output layer node

% Connect every node in the input layer with every node in the
% hidden layer.

%Input to hidden
\foreach \source in {1,...,\Nin}
\foreach \dest in {1,...,\Nhid}
\path (I-\source) edge (H-\dest);

\foreach \dest in {1,...,\Nhid}
\path (I-\Nin+1) edge (H-\dest);
\foreach \source in {1,...,\Nin}
\path (I-\source) edge (H-\Nhid+1);
\path (I-\Nin+1) edge (H-\Nhid+1);

%Hidden to hidden
\foreach \source in {1,...,\Nhid}
\foreach \dest in {1,...,\Nhidd}
\path (H-\source) edge (H2-\dest);

\foreach \dest in {1,...,\Nhidd}
\path (H-\Nhid+1) edge (H2-\dest);
\foreach \source in {1,...,\Nhid}
\path (H-\source) edge (H2-\Nhidd+1);
\path (H-\Nhid+1) edge (H2-\Nhidd+1);

%Hidden to output
\foreach \source in {1,...,\Nhidd}
\foreach \dest in {1,...,\Nout}
\path (H2-\source) edge (O-\dest);

\foreach \dest in {1,...,\Nout}
\path (H2-\Nhidd+1) edge (O-\dest);
\foreach \source in {1,...,\Nhid}
\path (H2-\source) edge (O-\Nout+1);
\foreach \source in {1,...,\Nhid}
\path (H2-\source) edge (O-\Nout+2);
\foreach \source in {1,...,\Nhid}
\path (H2-\source) edge (O-\Nout+3);
\path (H2-\Nhidd+1) edge (O-\Nout+1);
\path (H2-\Nhidd+1) edge (O-\Nout+2);
\path (H2-\Nhidd+1) edge (O-\Nout+3);

% Annotate the layers
\node[annot,above of=H-1, node distance=0.75cm] (hl) {Hidden layer};
\node[annot,above of=H2-1, node distance=0.75cm] (hl) {Hidden layer};
\node[annot,above of=I-1, node distance=0.75cm] {Input layer};
\node[annot,above of=O-1, node distance=0.75cm] {Output layer};

\node at (3\separ,-\Nout-1) {$\vdots$};
\end{tikzpicture}
	\caption{DNN to simultaneously forecast day-ahead prices in two markets.}
	\label{fig:hiddnet2}
	\end{center}
\end{figure}
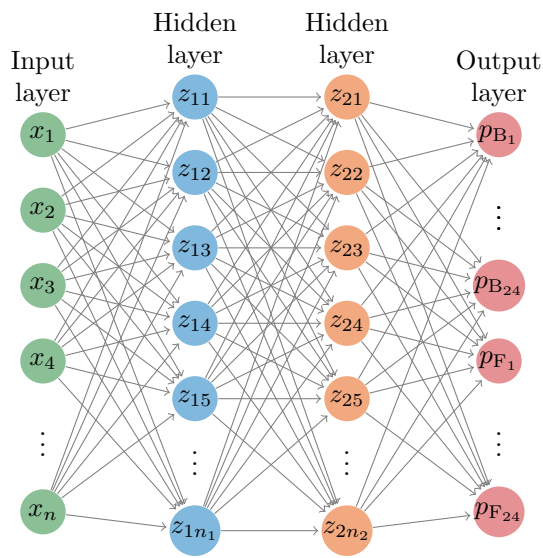

Consider an electricity market $\mathrm{B}$ and a second electricity market $\mathrm{F}$ that is connected to $\mathrm{B}$. Then, defining the output of the network by $\mathbf{p}=[p_{\mr{B}_{1}},\ldots,p_{\mr{B}_{24}},p_{\mr{F}_{1}},\ldots,p_{\mr{F}_{24}}]^\top\in\R^{48}$, i.e.~the set of 48 day-ahead prices from markets $\mr{B}$ and $\mr{F}$, and keeping the rest of the DNN parameter definitions the same, the new DNN structure can be represented as in Figure \ref{fig:hiddnet2}. In addition, as both models only differ in the output size, the implementation details are exactly the same as defined for the single-market model in Section \ref{sec:singledetails}.

\section{Feature Selection Algorithm}
\label{sec:feat}
As explained in the introduction, while the feature selection methods for electricity price forecasting proposed in the literature provide good and fast algorithms, they have two drawbacks:
\begin{enumerate}
	\item They perform a filter step where the model performance is not considered.
	\item For the nonlinear methods, the different inputs have to be transformed, i.e.~the selection is not done over the original feature set, and thus, some feature information might be lost.
\end{enumerate}

Therefore, we propose a nonlinear wrapper method that directly evaluates the features on the prediction model; in particular, while the approach is more computationally demanding, it can provide a better selection as it uses the real predictive performance without any data transformations. 

\subsection{Algorithm Definition}
In Section \ref{sec:hyper} we have introduced the TPE algorithm, a method for hyperparameter optimization, together with functional ANOVA, an approach for assessing hyperparameter importance. In this section, we combine both methods to build a feature selection algorithm that consists of four steps:
\begin{enumerate}
\item Model the features as hyperparameters.
\item Optimize the hyperparameters/features.
\item Analyze the results.
\item Select the important features.
\end{enumerate}

\subsubsection{Features as Hyperparameters}
The first step of the algorithm is to model the selection of features as model hyperparameters. In particular, we consider two types of features:

\begin{enumerate}
	\item Binary features $\theta_B$, whose selection can be done through a binary variable, i.e. $\theta_B\in\{0,1\}$,
	 where $\theta_B=0$ would represent feature exclusion and $\theta_B=1$ feature inclusion. Binary features represent the type of features considered by traditional algorithms. An example would be whether to include holidays data or whether to select a specific lag in an ARIMA model.
	\item Integer features $\theta_I$, which not only can model the {inclusion-exclusion} of an input, but also select some associated size or length, i.e. $\theta_I\in\mathbb{Z}$,
 where $\theta_I=0$ represents exclusion. Examples would be the number of past days of price data or the maximum lag of an ARIMA model.
\end{enumerate}

Given these definitions, the binary features are modeled as hyperparameters using the hyperparameter space $\vc{\Theta}_{B}$ and the hyperparameter set $B=\{1,\ldots,n_B\}$. Likewise, the integer features are modeled by the hyperparameter space $\vc{\Theta}_{I}$ and the hyperparameter set $I=\{n_B+1,\ldots,n_B+n_I\}$. Finally, the full hyperparameter space is defined by $\vc{\Theta}=\vc{\Theta}_{B}\cup\vc{\Theta}_{I}$ and the hyperparameter set by $\featset=B\cup I$.

%e.g.~considering a feature length $\theta_I$ only when a binary feature  $\theta_B$ is excluded, and to model inclusion-exclusion within the integer type,  $\theta_I=0$ can be used as exclusion, and then, by definition, $\theta_I>0$ would represent inclusion.

\subsubsection{Feature Optimization}
The second step of the algorithm is to perform a TPE optimization over the hyperparameter-feature space. The result of the algorithm is the optimal feature selection $\bm{\theta}^*$ together with the set $\mathcal{H}=\bigl\{(\bm{\theta}_\ixts,p_\ixts)\bigr\}_{\ixts=1}^\nruns$ of feature-performance pairs, where $p_\ixts$ represents the model predictive accuracy when using the feature selection $\bm{\theta}_\ixts$.

The fact that a feature is part of $\bm{\theta}^*$, does not guarantee that the feature is relevant; specifically, a feature might have little or no effect in the performance, and still, as long as it does not have a negative effect, it might appear in the optimal configuration. As a result, if no further processing is considered, the algorithm might select redundant features, and in turn, lead to more computationally expensive models and increase the risk of overfit. 

\subsubsection{Feature Analysis}
To solve the problem of detecting unnecessary features, the algorithm comprises a third step where feature importance is analyzed. In particular, using the functional ANOVA methodology proposed in \cite{Hutter2014}, the algorithm analyzes $\mathcal{H}$ and provides the importance of each feature $\ixset$ and each pairwise interaction $\{\ixset,j\}$ as the percentage-wise contribution to the performance variance $\mathbb{V}$.
% where $\ixset\neq j$ and $\Theta_\ixset\times \Theta_j$ represents the 2-dimensional state space of features $\ixset$ and $j$. 
Using the definitions given in Section \ref{sec:fanova} and \eqref{eq:variance}-\eqref{eq:importance}, the algorithm computes the importance of feature $\Theta_i$ and each pairwise interaction $\Theta_\ixset\times\Theta_j$ by:
\begin{equation}
\mathbb{F}_{\{\ixset\}} = \frac{\mathbb{V}_{\{\ixset\}}}{\mathbb{V}}, \quad \quad\quad\mathbb{F}_{\{\ixset,j\}}  = \frac{\mathbb{V}_{\{\ixset,j\}} }{\mathbb{V}}.
\end{equation}

In addition, for each feature $\ixset\in \featset$ and feature instantiation $ \theta_\ixset\in\Theta_\ixset$, the algorithm also provides the predicted marginal performance $\hat{a}(\theta_\ixset)$.

\subsubsection{Feature Selection}
The fourth and final algorithm step is the selection itself. In particular, making use of the obtained $\mathbb{F}_{\{\ixset\}}$, $\mathbb{F}_{\{\ixset,j\}}$ and $\hat{a}(\theta_\ixset )$, the selection procedure performs the following steps:
\begin{enumerate}
	\item Define a threshold parameter $\epsilon\in(0,1]$.
	\item Make a pre-selection by discarding features that do not improve nor decrease the performance. In particular, regard features $i$ whose importance $F_{\{\ixset\}}$ is larger than $\epsilon$:
	\begin{subequations}
		\begin{equation}
		U^*_1 =\{ \ixset \in \featset~|~F_{\{\ixset\}}>\epsilon \},
		\end{equation}
		or features $\ixset$ that have at least one pairwise contribution $F_{\{\ixset,j\}}$ larger than $\epsilon$:
		\begin{equation}
		U^*_{2} =\{\ixset \in \featset~|~\exists~j\in \featset\setminus\{\ixset\}:~F_{\{\ixset,j\}}>\epsilon\}.
		\end{equation}
		%\item Build the union of the previous sets:
		%\begin{equation}\[\end{equation}
		%\Theta^*=\Theta^*_1\cup\Theta^*_2
		%\]
		\item With the remaining features in $U^*_1\cup U^*_2$, perform a second selection $U^*$ by discarding those features whose predicted marginal performance $\hat{a}(\theta_\ixset)$ is lower when being included than when being excluded, i.e.:
		\begin{equation}
		U^* =\{ \ixset \in U^*_1\cup U^*_2~|~\exists~\theta_\ixset\in\Theta_\ixset:~ \mu_{\theta_{\ixset,0}} <\hat{a}(\theta_\ixset)\},
		\end{equation}
		\noindent where $\mu_{\theta_{\ixset,0}}$ represents the marginal performance $\hat{a}(\theta_\ixset=0)$ of excluding feature $\ixset$. 
		%This step eliminates features with a large contribution to the variance by having a better performance when being excluded, e.g.~by inducing overfitting.
		
		% This process is important because the previous step just discarded features with barely null contribution; therefore, some features might display a large contribution to the variance by having a better performance when being excluded, e.g.~by inducing overfitting.
		\item Finally, the set of selected binary features can be obtained by:
		\begin{equation}
		U^*_B = U^*\cap B.
		\end{equation}
		\noindent Similarly, for the set of optimal integer features $U^*_I$, the selection is done in terms of the feature itself and the instantiation with the best performance:
		\begin{alignat}{2}
		U^*_I =\bigl\{ &\{\ixset,\theta_\ixset^*\}~| ~\ixset \in U^*\cap I,~\theta_\ixset^* = \underset{{\theta_\ixset}}{\mr{argmax}}~ \hat{a}(\theta_\ixset) \bigr\}.
		\end{alignat}
	\end{subequations}
\end{enumerate}

\subsection{Case Study}
\label{subsec:featureE} 
To evaluate the proposed algorithm, we use it to select the features for predicting Belgian prices and to obtain a first assessment of the effect of market integration, i.e.~the effect of French features in forecasting Belgian prices. To perform the analysis, we consider the first and simpler DNN proposed in Section \ref{sec:modelframe}.

\subsubsection{Feature Definition}
In order to perform the feature selection, we first need to model each possible input as either a binary or an integer feature.
As described in Section \ref{sec:data}, the available features are
the day ahead prices $p_\mr{B}$ and $p_\mr{F}$, the day-ahead forecasts $l_\mr{B}$ and $l_\mr{F}$ of the grid load, the day-ahead forecasts  $g_\mr{B}$ and $g_\mr{F}$ of the available generation, and the calendar of public holidays $H_\mr{B}$ and $H_\mr{F}$. 

Considering that, given the market at time $\dt$, we aim at forecasting the time series vector $\mathbf{p}_{\mr{B}_\dt}=[p_{\mr{B}_{\dt+1}},\ldots,p_{\mr{B}_{\dt+24}}]^\top$ of Belgian day-ahead prices, the use of
the day-ahead loads $\mathbf{l}_{\mr{B}_\dt}=[l_{\mr{B}_{\dt+1}},\ldots,l_{\mr{B}_{\dt+24}}]^\top$ and $\mathbf{l}_{\mr{F}_\dt}=[l_{\mr{F}_{\dt+1}},\ldots,l_{\mr{F}_{\dt+24}}]^\top$, and the use of the day-ahead capacity generations $\mathbf{g}_{\mr{B}_\dt}=[g_{\mr{B}_{\dt+1}},\dots,g_{\mr{B}_{\dt+24}}]^\top$ and $\mathbf{g}_{\mr{F}_\dt}=[g_{\mr{F}_{\dt+1}},\dots,g_{\mr{F}_{\dt+24}}]^\top$, should be modeled as binary features $\theta_{l_\mr{B}}$, $\theta_{l_\mr{F}}$, $\theta_{g_\mr{B}}$, and $\theta_{g_\mr{F}}$. 
%In particular, if $\theta_{f_\mr{C}}$ is set to 1, where $f\in\{l,g\}$ and $C\in\{B,F\}$, the corresponding time series vector $\mathbf{X}_{f_{\mr{C},\dt}}=[f_{\mr{C}_{\dt+1}},\ldots,f_{\mr{C}_{\dt+24}}]^\top$ is used as a model input.

Similarly, for the public holidays, the features can also be modeled as binary variables $\theta_{H_\mr{B}}$ and $\theta_{H_\mr{F}}$. In particular, as the set of 24 hours of a day is either a holiday or not, the holidays are defined as model inputs $X_{H_\mr{B}},~X_{H_\mr{F}}\in\{0,1\}$, with $0$ and $1$ representing respectively no holiday and holiday. 
%Then, if either $\theta_{H_\mr{B}}$ or $\theta_{H_\mr{F}}$ is set to 1, the corresponding variable, i.e.~$X_{H_\mr{B}}$ or $X_{H_\mr{F}}$, is used as model input.

To model the Belgian prices, we need to use an integer feature to select the number of the considered past values. In particular, as the prices display daily and weekly seasonality, we have to use two integer features: $\theta_{p_\mr{B,d}}\in\{1,2,\ldots,6\}$ as the feature modeling the number of past days during the last week (daily seasonality) and $\theta_{p_\mr{B,w}}\in\{1,2,3\}$ as the feature modeling the number of days at weekly lags (weekly seasonality). Based on the selection of $\theta_{p_\mr{B,d}}$ and $\theta_{p_\mr{B,w}}$, the considered EPEX-Belgium past prices can be decomposed as the price inputs $\mathbf{X}^\mr{d}_{p_{\mr{B},h}}$ at daily lags and the price inputs $\mathbf{X}^\mr{w}_{p_{\mr{B},h}}$ at weekly lags:
\begin{subequations}
\begin{alignat}{2}
&\mathbf{X}^\mr{d}_{p_{\mr{B},\dt}}=\bigl[p_{\mr{B}_{h-i_1}}, \ldots,p_{\mr{B}_{h-i_{N_\mr{d}}}}\bigr]^\top,\\ 
&\mathbf{X}^\mr{w}_{p_{\mr{B},\dt}}=\bigl[p_{\mr{B}_{h-j_1}}, \ldots,p_{\mr{B}_{h-j_{N_\mr{w}}}}\bigr]^\top,
\end{alignat}
\noindent where:
\begin{alignat}{2}
&\{i_1,\ldots,i_{N_\mr{d}}\}&&=\{i~|~0\leq i\leq{24\cdot\theta_{p_\mr{B,d}}-1}\}\\
& \{j_1,\ldots,j_{N_\mr{w}}\}&&=\{j~|~1\leq k \leq \theta_{p_\mr{B,w}},\\
& && k\cdot 168\cdot\theta_{p_\mr{B,d}}\leq j \leq k\cdot 192\cdot\theta_{p_\mr{B,d}}-1\}\nonumber
\end{alignat}
\end{subequations}
 \noindent It is important to note that, as this is the time series to be predicted, we disregard the cases where no daily nor weekly seasonality is used, i.e.~$\theta_{p_\mr{B,d}}=0$ or $\theta_{p_\mr{B,w}}=0$.
 
 Finally, for the EPEX-France prices we could use the same integer features as for EPEX-Belgium. However, for simplicity, we directly consider the same lags for both time series and model the French prices as a binary feature $\theta_{p_\mr{F}}$. It is important to note that, despite having the same length, the selection of both time series is still independent; particularly, the lags are only defined for Belgium, and the French prices are just excluded or included. The modeled input features are summarized in Table \ref{tab:deffeat}.
 
  \begin{table}[h]
 	\small
 	\renewcommand\arraystretch{1.2}
 	\begin{center}
 		\begin{tabular}{|l|c|l|}
 			\hline
 			\bfseries Feature &\bfseries  Domain &\bfseries  Definition\\
 			\hline
 			$\theta_{p_\mr{B,d}}$&$\{1,\ldots,6\}$ & \specialcell[l]{\!\!\!\!\!\!\!\!\! Number of past days \\ for input price sequence }\\
 			\hline
 			$\theta_{p_\mr{B,w}}$&$\{1,\ldots,3\}$ & \specialcell[l]{\!\!\!\!\!\!\!\!\!\!\!Days at weekly  lags\\ for input price sequence}\\
 			\hline
 			$\theta_{p_\mr{F}}$&$\{0,1\}$ & Day-ahead price in France\\
 			\hline
 			$\theta_{l_\mr{B}}$&$\{0,1\}$& Load in Belgium\\
 			\hline
 			$\theta_{l_\mr{F}}$&$\{0,1\}$ & Load in France\\
 			\hline
 			$\theta_{g_\mr{B}}$&$\{0,1\}$ & Generation in Belgium\\
 			\hline
 			$\theta_{g_\mr{F}}$&$\{0,1\}$  & Generation in France\\
 			\hline
 			$\theta_{H_\mr{B}}$&$\{0,1\}$ & Holiday in Belgium\\
 			\hline
 			$\theta_{H_\mr{F}}$&$\{0,1\}$ & Holiday in France \\
 			\hline			
 		\end{tabular}
 	\end{center}
 	\caption{Definition of the modeled input features.}
 	\label{tab:deffeat}
 \end{table}

 \subsubsection{Hyperparameter Optimization}
 \label{sec:hyperopt}
 In order to guarantee that the network is adapted according to the input size, we simultaneously optimize the hyperparameters of the DNN, i.e.~the number of neurons $\nhidn_1$ and $\nhidn_2$. In particular, as the feature selection method is based on a hyperparameter optimization, we directly include the number of neurons as integer hyperparameters that are optimized together with the features. We set the domain of $\nhidn_1$ as the set of integers $\{100,101,\ldots,400\}$ and the one of $\nhidn_2$ as $\{0\}\cup\{48,49,\ldots,360\}$, where $\nhidn_2=0$ represents removing the second hidden layer and using a network of depth one.
 
 \subsubsection{Experimental Setup}
 In order to use the proposed algorithm, we first need to define the threshold $\epsilon$ for the minimum variance contribution; in our case, we select $\epsilon=0.5~\%$.  In addition, we also need to select the maximum number of iterations $\nruns$ of the TPE algorithm; we found $\nruns=1000$ to offer a good trade-off between performance and accuracy. Particularly, considering that training a single model takes \SI{2}{\minute}, the full feature selection requires \SI{30}{\hour}. While this might seem a long time, this step is only performed after some periodic time, e.g.~a month, to reassess feature dependencies; therefore, the proposed approach and settings yield a feasible and accurate method for the time scale of day-ahead prices.

For implementing the functional analysis of variance, we use the python library fANOVA developed by the authors of \cite{Hutter2014}. Likewise, for implementing the TPE algorithm, we use the python library hyperopt \cite{Bergstra2013}. 

\subsubsection{Results}
\label{sec:expresu1}
\begin{figure}[h]
	\begin{center}
		\input{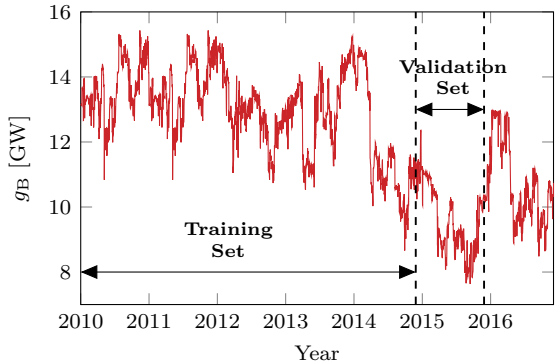}
		\caption{\scriptsize Generation in Belgium in the considered period.}
		\label{fig:fullgenbe}
	\end{center}
\end{figure}

In a first conducted experiment, we obtained an unexpected result: inclusion/exclusion of the generation capacity in Belgium $g_\mr{B}$ accounts for roughly $75\%$ of the performance variance $\mathbb{V}$, with inclusion of ${g_\mr{B}}$ dramatically decreasing the predictive accuracy. Since the generation capacity has been successfully used by other authors as a market driver \cite{Weron2014}, this result requires some explanation. From Figure \ref{fig:fullgenbe}, which displays the time series of ${g_\mr{B}}$, we can comprehend the result: right before the transition from the training to the validation set, the average $g_\mr{B}$ suffers a major change and drops from approximately \SI{14}{\giga\watt} to \SI{9}{\giga\watt}. Because of the drastic drop, it is likely that some relations that are learned based on the training set, do not hold in the validation set, and that as a result, the predictive performance in the validation set worsens when $g_\mr{B}$ is considered.

\setlength{\figW}{\columnwidth}
\setlength{\figH}{0.7\figW}	

This regime change in $g_\mr{B}$ violates the assumption that conditions in the training, validation, and test sets are equal. Therefore, to perform a correct feature selection and to guarantee that the three datasets hold similar conditions, the experimental setup should disregard $\theta_{g_\mr{B}}$. {It is important to note that, before taking this decision, we have considered shuffling the data to ensure homogeneous conditions between the three sets. However, this alternative was avoided for two reasons:
\begin{enumerate}
	\item As the output prices in some samples are the input features in others, data has to be discarded in order to avoid data contamination between the three sets. As a result, since the larger the dataset the better the DNN can generalize, this implementation could potentially decrease the predictive accuracy of the model.
	\item Since the end goal of the model is to forecast recent prices, it is meaningless to try to model an input-output relation that no longer holds.
\end{enumerate}}

Considering these facts, a correct feature selection is performed without $\theta_{g_\mr{B}}$. As depicted in Table \ref{tab:stdH}, the first result to be noted from the new experimental results is that, as $g_\mr{B}$ is a big source of error, the variance $\hat{\mathbb{V}}$ of the $\mape$ performance is reduced by a factor of 5. 
\begin{table}[htb]
	\small
	\renewcommand\arraystretch{1.2}
	\begin{center}
		\begin{tabular}{|l|c|}
			\hline
			& \textbf{$\hat{\mathbb{V}}$} \\
			\hline
			Feature selection with  $g_\mr{B}$ & 0.58  $\%^2$ \\
			\hline
			Feature selection without  $g_\mr{B}$ & 0.12 $\%^2$\\
			\hline
		\end{tabular}
	\end{center}
	\caption{Performance variance with and without $g_\mr{B}$.}
	\label{tab:stdH}
\end{table}
In addition, as it could be expected, the results obtained in this new experiment display a more distributed contribution among the different features. In particular, in the first experiment, $g_\mr{B}$ was responsible for 75\% of the performance variance. Now, as depicted in Table \ref{tab:verexp2}, French prices and load account for roughly 50 \%
of the total performance variance, and the available generation in France, the load in Belgium, and the number of past days play a minor role.

\begin{table}[htb]
	\small
	\renewcommand\arraystretch{1.2}
	\begin{center}
		\begin{tabular}{|l|c|}
			\hline
			& \textbf{Contribution to $\mathbb{V}$ } \\
			\hline
			All main effects & 64.9\%\\
			\hline
			French load & 28.4\%\\
			\hline
			French prices & 25.7\%\\			
			\hline
			French generation & 4.78\%\\			
			\hline
			Belgium load & 1.0\%\\			
			\hline		
			Past days number & 0.8\%\\			
			\hline											
		\end{tabular}
	\end{center}
	\caption{Variance contribution of single features for the second feature selection experiment.}
	\label{tab:verexp2}
\end{table}

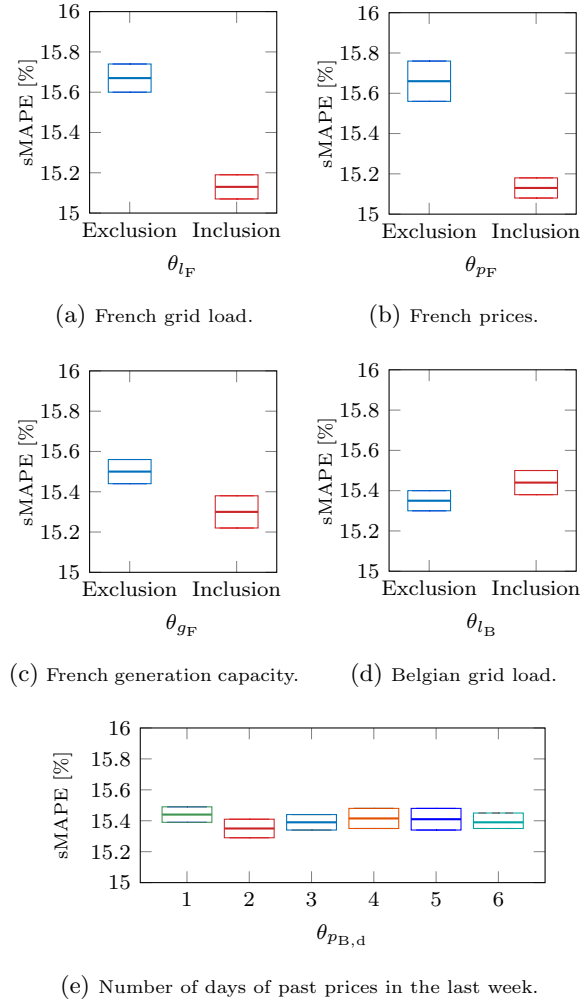
\begin{figure}[hbt]
	\setlength{\figW}{0.545\columnwidth}
	\setlength{\figH}{\figW}	
	\begin{subfigure}[t]{0.49\columnwidth}
		\begin{center}
			\begin{tikzpicture}
\begin{axis}[
width=0.951\figW,
height=\figH,    
boxplot/draw direction=y,
xtick={1,3},
xticklabels={Exclusion, Inclusion},  
ylabel={sMAPE}, 
y unit=\%,	   	
xlabel=$\theta_{l_\mr{F}}$,
xlabel style={font=\footnotesize},
ylabel style={font=\scriptsize},
ticklabel style={font=\footnotesize},
y label style={at={(axis description cs:-0.23,0.5)}},		
xmin=0.25,
xmax=3.75,
ymin=15,
ymax=16,			   	
]
\addplot+[boxplot prepared={
every box/.style={draw=bluePlots},
every median/.style={bluePlots, thick},      	      	      	      	      		
lower whisker=15.6, lower quartile=15.6,
median=15.67, upper quartile=15.74,
upper whisker=15.74}]
coordinates {};
\addplot+[boxplot prepared={draw position=3,
every box/.style={draw=redPlots},
every median/.style={redPlots, thick},      	      	      	      	      		
lower whisker=15.07, lower quartile=15.07,
median=15.13, upper quartile=15.19,
upper whisker=15.19}]
coordinates {};
\end{axis}
\end{tikzpicture}
			\caption{\scriptsize French grid load.}
		\end{center}
	\end{subfigure}
	\begin{subfigure}[t]{0.49\columnwidth}
		\begin{center}
			\begin{tikzpicture}
    \begin{axis}[
		width=0.951\figW,
		height=\figH,    
        boxplot/draw direction=y,
		xtick={1,3},
	    xticklabels={Exclusion, Inclusion},  
	   	ylabel={sMAPE}, 
y unit=\%,	   	
	   	xlabel=$\theta_{p_\mr{F}}$,
		xlabel style={font=\footnotesize},
		ylabel style={font=\scriptsize},
		ticklabel style={font=\footnotesize},
	    y label style={at={(axis description cs:-0.23,0.5)}},		
		xmin=0.25,
		xmax=3.75,
		ymin=15,
		ymax=16,			   	
      ]
      \addplot+[boxplot prepared={
      	every box/.style={draw=bluePlots},
every median/.style={bluePlots, thick},      	      	      	      	      	      	
          lower whisker=15.56, lower quartile=15.56,
		median=15.66, upper quartile=15.76,
		upper whisker=15.76}]
          coordinates {};
      \addplot+[boxplot prepared={draw position=3,
      	every box/.style={draw=redPlots},
every median/.style={redPlots, thick},      	      	      	      	      	      	
          lower whisker=15.08, lower quartile=15.08,
          median=15.13, upper quartile=15.18,
          upper whisker=15.18}]
          coordinates {};
    \end{axis}
  \end{tikzpicture}
			\caption{\scriptsize French prices.}
		\end{center}
	\end{subfigure} \\[0.3cm]
	\begin{subfigure}[t]{0.49\columnwidth}
		\centering
		\begin{tikzpicture}
    \begin{axis}[
		width=0.951\figW,
		height=\figH,    
        boxplot/draw direction=y,
		xtick={1,3},
	    xticklabels={Exclusion, Inclusion},  
	   	ylabel={sMAPE}, 
y unit=\%,	   	
	   		   	xlabel=$\theta_{g_\mr{F}}$,
		xlabel style={font=\footnotesize},
		ylabel style={font=\scriptsize},
		ticklabel style={font=\footnotesize},
	    y label style={at={(axis description cs:-0.23,0.5)}},		
		xmin=0.25,
		xmax=3.75,
		ymin=15,
		ymax=16,	
      ]
      \addplot+[boxplot prepared={
      	every box/.style={draw=bluePlots},
every median/.style={bluePlots, thick},      	      	      	      	      	      	
          lower whisker=15.44, lower quartile=15.44,
		median=15.50, upper quartile=15.56,
		upper whisker=15.56}]
          coordinates {};
      \addplot+[boxplot prepared={draw position=3,
      	every box/.style={draw=redPlots},
every median/.style={redPlots, thick},      	      	      	      	      	      	
          lower whisker=15.22, lower quartile=15.22,
          median=15.30, upper quartile=15.38,
          upper whisker=15.38}]
          coordinates {};
    \end{axis}

  \end{tikzpicture}
		\caption{\scriptsize French generation capacity.}
	\end{subfigure} 
	\begin{subfigure}[t]{0.49\columnwidth}
		\centering
		\begin{tikzpicture}
    \begin{axis}[
		width=0.951\figW,
		height=\figH,    
        boxplot/draw direction=y,
		xtick={1,3},
	    xticklabels={Exclusion, Inclusion},  
	   	ylabel={sMAPE}, 
y unit=\%,	   	
		xlabel=$\theta_{l_\mr{B}}$,	   	
		xlabel style={font=\footnotesize},
		ylabel style={font=\scriptsize},
		ticklabel style={font=\footnotesize},
	    y label style={at={(axis description cs:-0.23,0.5)}},		
		xmin=0.25,
		xmax=3.75,
		ymin=15,
		ymax=16,	
      ]
      \addplot+[boxplot prepared={
      	every box/.style={draw=bluePlots},
every median/.style={bluePlots, thick},      	      	      	      	      	      	
          lower whisker=15.3, lower quartile=15.3,
		median=15.35, upper quartile=15.4,
		upper whisker=15.4}]
          coordinates {};
      \addplot+[boxplot prepared={draw position=3,
      	      	every box/.style={draw=redPlots},
      	every median/.style={redPlots, thick},      	      	      	      	      	
          lower whisker=15.38, lower quartile=15.38,
          median=15.44, upper quartile=15.50,
          upper whisker=15.50}]
          coordinates {};
    \end{axis}
  \end{tikzpicture}
		\caption{\scriptsize Belgian grid load.}
	\end{subfigure} 
	\\[0.3cm]
	\begin{subfigure}[t]{0.98\columnwidth}
		\setlength{\figW}{0.95\columnwidth}
		\setlength{\figH}{0.5\figW}			
		\centering
		\begin{tikzpicture}
    \begin{axis}[
		width=0.951\figW,
		height=\figH,    
        boxplot/draw direction=y,
		xtick={1,2,3,4,5,6},
	   	ylabel={sMAPE}, 
y unit=\%,	   	
	   		   	xlabel=$\theta_{p_\mr{B,d}}$,
		xlabel style={font=\footnotesize},
		ylabel style={font=\scriptsize},
		ticklabel style={font=\footnotesize},
		xmin=0.25,
		xmax=6.75,
		ymin=15,
		ymax=16,	
      ]
      \addplot+[boxplot prepared={
      	every box/.style={draw=greenPlots},
      	every median/.style={greenPlots, thick},			      		
          lower whisker=15.39, lower quartile=15.39,
		median=15.44, upper quartile=15.49,
		upper whisker=15.49}]
          coordinates {};
      \addplot+[boxplot prepared={
      	every box/.style={draw=redPlots},
      	every median/.style={redPlots, thick},      	    	
          lower whisker=15.29, lower quartile=15.29,
          median=15.35, upper quartile=15.41,
          upper whisker=15.41}]
          coordinates {};
      \addplot+[boxplot prepared={
      	every box/.style={draw=bluePlots},
      	every median/.style={bluePlots, thick},      	      	
		lower whisker=15.34, lower quartile=15.34,
		median=15.39, upper quartile=15.44,
		upper whisker=15.44}]
		coordinates {};     
      \addplot+[boxplot prepared={
      	every box/.style={draw=orange},
		every median/.style={orange, thick},      	      	      	
	lower whisker=15.35, lower quartile=15.35,
	median=15.415, upper quartile=15.48,
	upper whisker=15.48}]
coordinates {};          		     
      \addplot+[boxplot prepared={
      	every box/.style={draw=blue},
		every median/.style={blue, thick},      	      	      	      	
	lower whisker=15.34, lower quartile=15.34,
	median=15.41, upper quartile=15.48,
	upper whisker=15.48}]
coordinates {};    
      \addplot+[boxplot prepared={
      	every box/.style={solid,draw=turqoise},
		every median/.style={turqoise, solid, thick},      	      	      	      	      	
	lower whisker=15.35, lower quartile=15.35,
	median=15.39, upper quartile=15.45,
	upper whisker=15.45}]
coordinates {};          		     
    \end{axis}
  \end{tikzpicture}
		\caption{\scriptsize Number of days of past prices in the last week.}
	\end{subfigure}
	%	\begin{subfigure}[t]{0.49\columnwidth}
	%	\centering
	%	\input{figs/neurons2}
	%	\caption{\scriptsize Number of neurons of second hidden layer.}
	%\end{subfigure}
	\caption{ Marginal performance on the validation set of the five most important features.}
	\label{fig:fr}
\end{figure}

Based on the above results, we can make a first selection and remove from the set of possible inputs the public holidays $\theta_{H_\mr{B}}$ and $\theta_{H_\mr{F}}$ as both seem not to be decisive. Similarly, we can select $\theta_{p_\mr{B,w}}=1$ as the number of days at weekly lags seems to be non-critical. Finally, to complete the feature selection, we should use the marginal performances of the five important features represented in Figure \ref{fig:fr}; based on them, it is clear that we should select the price, load and generation in France, discard the grid load in Belgium, and use two days of past price data.

Together with the features, we have also optimized the hyperparameters of the model. The results show that the suitable numbers of neurons are $\nhidn_2=200$ and $\nhidn_1=320$.

\subsection{Discussion}
\label{sec:disc5}
Based on the results of the feature selection algorithm, we should include the following features as model inputs:
\begin{enumerate}
\item Day-ahead load and generation in France:
%\begin{subequations}
%\begin{alignat}{2}
%\mathbf{X}_{l_{\mr{F},\dt}}&=[l_{\mr{F}_{\dt+1}},\,l_{\mr{F}_{\dt+2}},\ldots,\,l_{\mr{F}_{\dt+24}}]^\top,\\
%\mathbf{X}_{g_{\mr{F},\dt}}&=[g_{\mr{F}_{\dt+1}},\,g_{\mr{F}_{\dt+2}},\dots,\,g_{\mr{F}_{\dt+24}}]^\top.
%\end{alignat}
\item Last two days of Belgian and French prices:
%\begin{alignat}{2}
%\mathbf{X}^\mr{d}_{p_{\mr{B},\dt}}=[p_{\mr{B}_{\dt-0}},\,p_{\mr{B}_{\dt-1}},\ldots,\,p_{\mr{B}_{\dt-47}}]^\top,\\ % ~|~ &\ixset\in[0,47]\},\\
%\mathbf{X}^\mr{d}_{p_{\mr{F},\dt}}=[p_{\mr{F}_{\dt-0}},\,p_{\mr{F}_{\dt-1}},\ldots,\,p_{\mr{F}_{\dt-47}}]^\top. % ~|~ &\ixset\in[0,47]\},\\
%%\mathbf{X}^\mr{d}_{p_{\mr{F},\dt}}=\{p_{\mr{F}_{\dt-\ixset}} ~|~ &\ixset\in[0,47]\}.
%\end{alignat}
\item Belgian and French prices a week before:
%\begin{alignat}{2}
%\!\!\!\!\!\mathbf{X}^\mr{w}_{p_{\mr{B},\dt}}=[p_{\mr{B}_{\dt-168}},\,p_{\mr{B}_{\dt-169}},\ldots,\, p_{\mr{B}_{\dt-191}}]^\top,\\
%\!\!\!\!\! \mathbf{X}^\mr{w}_{p_{\mr{F},\dt}}=[p_{\mr{F}_{\dt-168}},\,p_{\mr{F}_{\dt-169}},\ldots,\,p_{\mr{F}_{\dt-191}}]^\top.
%\end{alignat}
%\end{subequations}
\end{enumerate}

 In addition, while it seems that the different French market features, i.e.~market integration features, play a large role in the forecasting accuracy, the results are only enough to have a general idea of the importance of French data; particularly, a statistical analysis is required before making any further conclusion.

Finally, while we have used the proposed algorithm to select the input features, we have not yet provided an evaluation of its accuracy. In particular, to assess  its performance, we could compare models using only optimally selected features against models using also features that have been discarded; more specifically, we could evaluate the difference in predictive accuracy by means of hypothesis testing (see Section \ref{sec:accfeat}). 
%Since hypothesis testing and results for different models are already presented in the next section, we leave this evaluation for Section \ref{sec:accfeat}.

\section{Evaluation of Market Integration and Modeling Framework}
\label{sec:results}
%As stated through the different parts of the paper, the main goal of this research is to propose different models for including market integration in day-ahead forecasting, and then, to illustrate how market integration improves the predictive accuracy of price forecasting. While the models are already proposed in Section \ref{sec:modelframe}, and while we are able to use the results of the feature selection to obtain a first qualitative assessment of the effect of market integration, a deeper study is still required.

The analysis provided by the feature selection algorithm is based on the validation set; while this dataset  is not used for training the network, it is employed for early stopping and hyperparameter optimization. Therefore, to have a fully fair and unbiased evaluation, we need an extra comparison using unseen data to the full training process. Moreover, as the feature selection results were obtained using the first proposed model, results for the second model are also required. Finally, to have a meaningful assessment, the statistical significance of the results should be computed. To fulfill the requirements, the goal of this section is twofold:

\begin{enumerate}
\item Provide statistical significance of the improvements of using French market data, i.e.~market integration, by performing a DM test on the out-of-sample data represented by the test set.
\item Based on the same statistical test, demonstrate how a dual-market forecaster can provide significant improvements in predictive accuracy.
\end{enumerate}

%The section is organized as follows: Section \ref{sec:DBprac} describes the DM test for our case study. Then, Sections \ref{sec:62} and \ref{sec:63} evaluate the proposed models and compute the statistical significance of their improvements. Finally, Section \ref{sec:65} discusses the obtained results and Section \ref{sec:66} introduces several practical applications.

\subsection{Diebold-Mariano Test}
\label{sec:DBprac}
To assess the statistical significance in the difference of predictive accuracy, we use the DM test as defined by \eqref{eq:lossdif}-\eqref{eq:lf}.
 Since the neural network is trained using the absolute mean error, we choose to use also the absolute error to build the loss differential:
 \begin{equation}
 d^{M_1,M_2}_k = |\varepsilon_\ixts^{M_1}| - |\varepsilon_\ixts^{M_2}|.
 \end{equation}

In addition, we follow the same procedure as in \cite{Ziel2015} and we perform an independent DM test for each of the 24 time series representing the different hours of a day. The reason for this is that, as we use the same information to forecast the set of 24 prices, the forecast errors within the same day would exhibit a high correlation. Moreover, to have an assessment of the whole error sequence, we also perform the DM test considering serial correlation of order $k$ in the error sequence. Particularly, recalling that optimal $k$-step-ahead forecast errors are at most $(k-1)$-dependent \cite{Diebold1995}, we perform a DM test on the full loss differential considering serial correlation of order $23$.

In the various experimental setups of this case study, we employ the one-sided DM test given by \eqref{eq:osdm} at the 95\% confidence level.
%, i.e.~we test the null hypothesis of a forecaster A having the same or worse accuracy than a forecaster B against the alternative hypothesis of the forecaster A having better accuracy. 
This selection is done because we want to assess whether the performance of a forecaster A is statistically significantly better than a forecaster B, not whether the performances of forecasters A and B are significantly different (like it would be the case in the two-sided DM test). In more detail, for each hour $h=1,\dots,24$ of the day, we test the null hypothesis of a model $M_{\mr{1}}$ that uses French data having the same or worse accuracy than a model $M_{\mr{2}}$ that uses no French data. More specifically, we perform the following tests:

\begin{equation}
\begin{cases}
H_0:~\mathbb{E}({d}^{M_{\mr{1}},M_{\mr{2}}}_{h_k})\geq0,\\
H_1:~\mathbb{E}({d}^{M_{\mr{1}},M_{\mr{2}}}_{h_k})<0,
\end{cases}~\mr{for}~h=1,\ldots24,
\end{equation}

\noindent where $[{d}_{h_1},\ldots,{d}_{h_{N/24}}]^\top$ represents the vector sequence of loss differentials of hour $h$. In addition, we perform the same test but considering the full loss differential sequence and assuming serial correlation:

\begin{equation}
\begin{cases}
H_0:~\mathbb{E}({d}^{M_{\mr{1}},M_{\mr{2}}}_{k})\geq0,\\
H_1:~\mathbb{E}({d}^{M_{\mr{1}},M_{\mr{2}}}_{k})<0.
\end{cases}
\end{equation}

\subsection{French Market Data: Statistical Significance}
\label{sec:62}
In Section \ref{subsec:featureE}-\ref{sec:disc5}, we have showed that using market data from connected markets can help to improve the performance. In this section, we extend the analysis by directly comparing a model that includes this type of data against a model that excludes it, and then, performing a DM test to analyze the statistical significance.

\subsubsection{Experimental Setup}
The model used to perform the evaluation is the single-market forecaster employed for the feature selection. In particular, based on the obtained hyperparameter results, we select $\nhidn_1=320$ and $\nhidn_2=200$; similarly, considering the optimized prices lags obtained in the feature selection, we consider, as input sequence for the model, the Belgium prices during the last two days and a week before. Then, we discard as input features the capacity generation in Belgium as well as the holidays in both countries. Then, in order to compare the effect of French data, we consider the remaining features as possible inputs for the model, i.e.~we compare the first model excluding all the French data and only considering Belgian prices with respect to the second model including the French data. We respectively refer to these two models as $M_\mr{NoFR}$ and $M_\mr{FR}$.

In addition, while the load in Belgium $l_\mr{B}$ appears to be non-relevant, we decided to repeat the previous experiment but including $l_\mr{B}$ in both models. The reason for this is twofold: 
\begin{enumerate}
\item By adding the Belgian load, we ensure that the good results of using French data are not due to the fact that the model does not include specific Belgian regressors.
\item Furthermore, with this experiment, we can also validate the results of the feature selection algorithm. In particular, as the load does not seem to play a big role, we expect the performance difference between models with and without $l_\mr{B}$ to be insignificant.
%\item By generating an extra experiment, the experimental setup size is increased and the obtained results are more reliable.
\end{enumerate}
\noindent Similar as before, we refer to these models by $M_{\mr{NoFR},l_\mr{B}}$ and $M_{\mr{FR},l_\mr{B}}$.

\subsubsection{Case 1: Models Without $l_\mr{B}$}
\label{sec:result11}
In this experiment, we compare  $M_\mr{NoFR}$ against $M_\mr{FR}$ by evaluating their performance on the year of yet unused data represented by the test set. As in a real-world application, to account for the last available information, the two models are re-estimated after a number days/weeks. In our application, considering that a model takes around 2 minutes to be trained on the GPU, we decide to re-estimate them using the smallest possible period of a day.

A first comparison of the models is listed in Table \ref{tab:ex1smape} by means of $\mape$. From this first evaluation, we can see that including the French data seems to really enhance the performance of the forecaster.

\begin{table}[h]
	\small
	\renewcommand\arraystretch{1.2}
	\begin{center}
		\begin{tabular}{|l|c|c|c|c|}
			\hline
			\bfseries Model &  $M_\mr{NoFR}$ & $M_\mr{FR}$ \\
			\hline
			$\mathrm{\mathbf{sMAPE}}$ & $16.0\%$&$13.2\%$\\
			\hline
		\end{tabular}
	\end{center}
	\caption{Performance comparison between $M_\mr{NoFR}$ and $M_\mr{FR}$ in the out-of-sample data in terms of $\mape$.}
	\label{tab:ex1smape}
\end{table}

To provide statistical significance to the above result, we perform a DM test as described in Section \ref{sec:DBprac}. The obtained results are depicted in Figure \ref{fig:exDM}, where the test statistic is represented for each of the 24 hours of a day and where the points above the dashed line accept, with a 95 \% confidence level, the alternative hypothesis of $M_\mr{FR}$ having better performance accuracy. As we can see from the plot, the forecast improvements of the model $M_\mr{FR}$ including French data are statistically significant for each one of the 24 day-ahead prices. 
\newcommand{\expnumber}[2]{${#1}\cdot{10}^{#2}$}
\setlength{\figW}{1.14\columnwidth}
\setlength{\figH}{0.4\figW}	
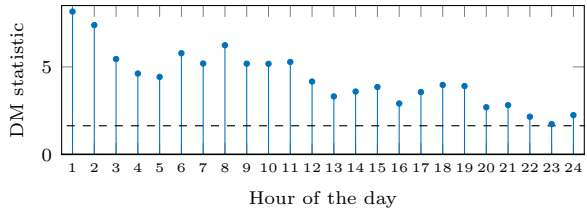
\begin{figure}[h]
	\begin{center}
		\begin{tikzpicture} 
\begin{axis}[
width=0.951\figW,
height=\figH, 
ylabel={DM statistic},
xlabel={Hour of the day},
xlabel style={font=\scriptsize},
ylabel style={font=\scriptsize},
ticklabel style={font=\scriptsize},
xticklabel style={font=\tiny},
xmin=0.5,
xmax=24.5,
ymin=0,
ymax=8.5,	
ytick={-5,0,5},
xtick={1,2,3,4,5,6,7,8,9,10,11,12,13,14,15,16,17,18,19,20,21,22,23,24},
]
\addplot+[ycomb,bluePlots,mark=*,mark options={scale=0.5, fill=bluePlots},] 
table {%
	1 8.1576185883066
	2 7.38588577561323
	3 5.44828651165273
	4 4.62173289980412
	5 4.42975916243917
	6 5.78015707463158
	7 5.19328877184455
	8 6.23555348306612
	9 5.18502408857699
	10 5.17813509913709
	11 5.27971434101224
	12 4.16521159027415
	13 3.31823365063598
	14 3.59643796060059
	15 3.85281152454686
	16 2.91357950167322
	17 3.56580442449903
	18 3.97358016732966
	19 3.90891537126712
	20 2.69999080705827
	21 2.82159572112581
	22 2.16071457072498
	23 1.74043165505808
	24 2.24946433633688
};
%plot coordinates {
%	(0,3) 
%	(1,2) 
%	(2,4) 
%	(3,1) 
%	(4,1)
%	(5,1)
%	(6,1)
%	(7,1)
%	(8,1)
%	(9,1)
%	(10,1)
%	(11,1)
%	(12,1)
%	(13,1)
%	(14,1)
%	(15,1)
%	(16,1)
%	(17,1)
%	(18,1)
%	(19,1)
%	(20,1)
%	(21,1)
%	(22,1)
%	(23,1)
%	(24,1)};
	
\addplot [color=black,line width=0.15mm,dashed,forget plot]
table[row sep=crcr]{%
	0 1.6448536269514722\\
	25 1.6448536269514722\\
};	

\addplot [color=black,line width=0.65mm,thick,forget plot]
table[row sep=crcr]{%
	0 0\\
	25 0\\
};	
	
\end{axis} 
\end{tikzpicture}
		\vspace{-0.5cm}
		\caption{ DM test results when comparing $M_\mr{NoFR}$ and $M_\mr{FR}$. Values above the dashed line reject the null hypothesis with a 95 \% confidence level, and in turn, represent cases where the accuracy of $M_\mr{FR}$ is significantly better.}
		\label{fig:exDM}
	\end{center}
\end{figure}

When the DM test is performed on the full loss differential and taking into account serial correlation, the obtained metrics completely agree with the results obtained for the individual 24 hours. In particular, the obtained $p$-value is \expnumber{1.2}{-11}, which confirms the strong statistical significance of using the French data in the prediction model.

%\begin{table}[h]
%	\small
%	\renewcommand\arraystretch{1.2}
%	\begin{center}
%		\begin{tabular}{|l|c|c|c|c|}
%			\hline
%			 $p$-value & DM Statistic \\
%			\hline
%			\expnumber{1.2}{-11} &6.67\\
%			\hline
%		\end{tabular}
%	\end{center}
%	\caption{$p$-value and statistic for the DM test on the full loss differential between $M_\mr{NoFR}$ and $M_\mr{FR}$.}
%	\label{tab:ex1fullDM}
%\end{table}

\subsubsection{Case 2: Models with $l_\mr{B}$}
Using the same procedure, we compare $M_{\mr{NoFR},l_\mr{B}}$ against $M_{\mr{FR},l_\mr{B}}$. From Table \ref{tab:ex2smape} we can see how, as before, the model including French data outperforms the alternative.

\begin{table}[h]
	\small
	\renewcommand\arraystretch{1.2}
	\begin{center}
		\begin{tabular}{|l|c|c|c|c|}
			\hline
			\bfseries Model &  $M_{\mr{NoFR},l_\mr{B}}$ & $M_{\mr{FR},l_\mr{B}}$ \\
			\hline
			$\mathrm{\mathbf{sMAPE}}$ & $15.7\%$&$13.1\%$\\
			\hline
		\end{tabular}
	\end{center}
	\caption{Performance comparison between $M_{\mr{NoFR},l_\mr{B}}$  and $M_{\mr{FR},l_\mr{B}}$ in the out-of-sample data in terms of $\mape$.}
	\label{tab:ex2smape}
\end{table}

\begin{figure}[h]
\setlength{\figW}{1.14\columnwidth}
\setlength{\figH}{0.4\figW}		
	\begin{center}
		\begin{tikzpicture} 
\begin{axis}[
width=0.951\figW,
height=\figH, 
ylabel={DM statistic},
xlabel={Hour of the day},
xlabel style={font=\scriptsize},
ylabel style={font=\scriptsize},
ticklabel style={font=\scriptsize},
xticklabel style={font=\tiny},
xmin=0.5,
xmax=24.5,
ymin=0,
ymax=8.5,	
ytick={-5,0,5},
xtick={1,2,3,4,5,6,7,8,9,10,11,12,13,14,15,16,17,18,19,20,21,22,23,24},
]
\addplot+[ycomb,bluePlots,mark=*,mark options={scale=0.5, fill=bluePlots},] 
table {%
1     6.494232
2     7.089215
3     6.139261
4     5.074631
5     3.818419
6     4.047253
7     5.253325
8     4.126376
9     3.446988
10    3.764000
11    4.492795
12    4.445868
13    4.285530
14    4.364664
15    4.918456
16    2.560846
17    3.493916
18    3.060469
19    3.691479
20    3.481316
21    4.676450
22    3.074179
23    3.185754
24    3.942067
};
%plot coordinates {
%	(0,3) 
%	(1,2) 
%	(2,4) 
%	(3,1) 
%	(4,1)
%	(5,1)
%	(6,1)
%	(7,1)
%	(8,1)
%	(9,1)
%	(10,1)
%	(11,1)
%	(12,1)
%	(13,1)
%	(14,1)
%	(15,1)
%	(16,1)
%	(17,1)
%	(18,1)
%	(19,1)
%	(20,1)
%	(21,1)
%	(22,1)
%	(23,1)
%	(24,1)};
	
\addplot [color=black,line width=0.15mm,dashed,forget plot]
table[row sep=crcr]{%
	0 1.6448536269514722\\
	25 1.6448536269514722\\
};	

\addplot [color=black,line width=0.65mm,thick,forget plot]
table[row sep=crcr]{%
	0 0\\
	25 0\\
};	
	
\end{axis} 
\end{tikzpicture}
		\vspace{-0.5cm}
		\caption{ DM test results when comparing $M_{\mr{NoFR},l_\mr{B}}$ and $M_{\mr{FR},l_\mr{B}}$. Values above the dashed line reject the null hypothesis at a 5\% significance level, and in turn, represent cases where the accuracy of $M_{\mr{FR},l_\mr{B}}$ is significantly better.}
		\label{fig:ex2DM}
	\end{center}
\end{figure}
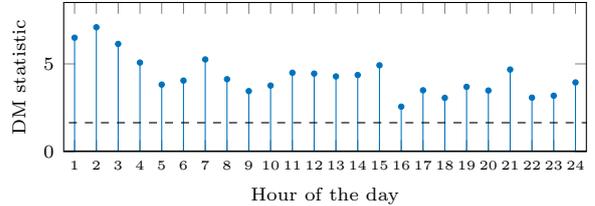

To provide statistical significance to the obtained accuracy difference we again perform the DM tests. The obtained results are illustrated in Figure \ref{fig:ex2DM}; as before, including French data leads to improvements in accuracy that are statistically significant for the 24 predicted values.
As before, when we consider the DM test for the full loss differential with serial correlation, the $p$-value is \expnumber{1.6}{-12}, a value that agrees with Figure \ref{fig:ex2DM} and confirms once more that the improvements of using French data are statistically significant.

%\begin{table}[h]
%	\small
%	\renewcommand\arraystretch{1.2}
%	\begin{center}
%		\begin{tabular}{|l|c|c|c|c|}
%			\hline
%			$p$-value & DM Statistic \\
%			\hline
%			\expnumber{1.6}{-12} &6.97\\
%			\hline
%		\end{tabular}
%	\end{center}
%	\caption{$p$-value and DM statistic for the DM test comparing the full error sequence with error correlation between $M_\mr{NoFR}$ and $M_\mr{FR}$.}
%	\label{tab:ex1fullDM}
%\end{table}

\subsubsection{Accuracy of the Feature Selection}
\label{sec:accfeat}
Using the results of the previous two sections, we can illustrate the accuracy of the proposed feature selection algorithm in Section \ref{sec:feat}. In particular, when performing the feature selection, we have observed that the contribution of the Belgian load $l_\mr{B}$ was rather insignificant and even slightly negative; this led to discard $l_\mr{B}$ as an input feature. 
In this section, to verify that the selection algorithm performed the right choice, we perform DM tests to compare $M_{\mr{NoFR},l_\mr{B}}$ against $M_{\mr{NoFR}}$ and $M_{\mr{FR},l_\mr{B}}$ against $M_{\mr{FR}}$. In particular, we perform a two-sided DM test per model pair with the null hypothesis of the models having equal accuracy. 

For the sake of simplicity, we avoid depicting the DM test results for each individual hour; instead we directly illustrate the $p$-values of the DM test when considering the whole loss differential sequence with serial correlation. As can be seen from Table \ref{tab:exload}, the obtained $p$-values for both tests are above 0.05, and as a result, the null hypothesis of equal accuracy cannot be rejected, i.e.~there is no statistical evidence of the models using Belgian load having different accuracy than the models without it.

\begin{table}[h]
	\small
	\renewcommand\arraystretch{1.4}
	\begin{center}
		\begin{tabular}{|l|c|c|c|c|}
			\hline
			\textbf{Model Pair} & $p$-\textbf{value} \\
			\hline				
$M_{\mr{FR},l_\mr{B}}$ vs $M_{\mr{FR}}$ & 0.435 \\			
			\hline
			$M_{\mr{NoFR},l_\mr{B}}$ vs $M_{\mr{NoFR}}$ & 0.275 \\
			\hline
		\end{tabular}
	\end{center}
	\caption{$p$-values for DM test with the null hypothesis of models with $l_\mr{B}$ having equal accuracy as models without it.}
	\label{tab:exload}
\end{table}

Based on the obtained results, it is clear that using $l_\mr{B}$ is not relevant, and thus, that the choice performed by the feature selection algorithm is correct. In particular, while this experiment does not analyze the performance of the feature selection on all the inputs, it does consider the most problematic feature. More specifically, as many researchers have successfully used the load as an explanatory variable \cite{Misiorek2006,Nogales2002,Cruz2011,Amjady2009,Panapakidis2016} and as the load itself does not display any regime change in the considered time interval, it is rather striking to see its minimal effect on the performance. Therefore, by demonstrating that the algorithm is correct when discarding the load, we obtain an assessment of its general accuracy, and we can conclude that the algorithm performs a correct feature selection.

\subsection{Evaluation of a Dual-Market Forecaster}
\label{sec:63}
%So far, all the illustrated results have considered the first model described in Section \ref{sec:modelframe}, i.e.~the single-market forecaster. However, in the same section, we have also proposed a second model which forecasted in a single model the day-ahead prices of Belgium and France. The idea of it was that, by solving closely related tasks, the neural network could learn more general features and avoid overfitting.

In this section, we evaluate the possible improvements of using the dual-market forecaster and multi-tasking by comparing the single-market model against the dual-market forecaster predicting the day-ahead prices in Belgium and France. The models are denoted by $M_{\mr{Single}}$ and $M_{\mr{{Dual}}}$
%,where the subindex $\mr{S}$ and $\mr{D}$ represents the single and dual forecaster, 
and they both use the optimal features and hyperparameters obtained for the single-market model in Section \ref{sec:feat}. 
It is important to note that, while in an ideal experiment the hyperparameters of the dual-market forecaster should be re-estimated, for simplicity we decided to directly use the hyperparameters obtained for the single-market forecaster.

The initial comparison is listed in Table \ref{tab:ex3smape}. From this first evaluation it seems that using dual-market forecasts can improve the performance.

\begin{table}[h]
	\small
	\renewcommand\arraystretch{1.2}
	\begin{center}
		\begin{tabular}{|l|c|c|c|c|}
			\hline
			\bfseries Model &  $M_{\mr{Single}}$ & $M_{\mr{{Dual}}}$ \\
			\hline
			$\mathrm{\mathbf{sMAPE}}$ & $13.2\%$&$12.5\%$\\
			\hline
		\end{tabular}
	\end{center}
	\caption{Performance comparison between the single and dual-market forecasters in terms of $\mape$.}
	\label{tab:ex3smape}
\end{table}

To provide statistical significance to these results, we again perform the DM test for each of the 24 hours of a day. The obtained statistics are depicted in Figure \ref{fig:ex3DM}; as before, the points above the upper dashed line accept, with a 95 \% confidence level, the alternative hypothesis of $M_{\mr{{Dual}}}$ having a better performance accuracy. In addition, as not every hourly forecast is statistically significant, we represent in the same figure the alternative DM test with the null hypothesis of $M_{\mr{{Single}}}$ having equal or lower accuracy than $M_{\mr{{Dual}}}$. This test is characterized by the lower dashed line and any point below this line accepts, with a 95 \% confidence level, that $M_{\mr{{Single}}}$ has better performance accuracy.

\setlength{\figW}{1.11\columnwidth}
\setlength{\figH}{0.4\figW}	
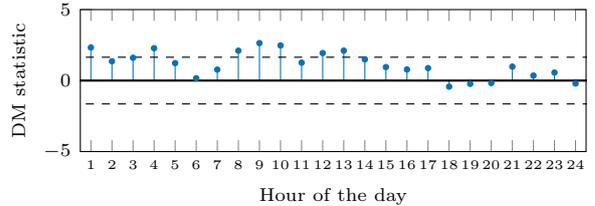
\begin{figure}[h]
	\begin{center}
		\begin{tikzpicture} 
\begin{axis}[
width=0.951\figW,
height=\figH, 
ylabel={DM statistic},
xlabel={Hour of the day},
xlabel style={font=\scriptsize},
ylabel style={font=\scriptsize},
ticklabel style={font=\scriptsize},
xticklabel style={font=\tiny},
xmin=0.5,
xmax=24.5,
ymin=-5,
ymax=5,	
ytick={-5,0,5},
xtick={1,2,3,4,5,6,7,8,9,10,11,12,13,14,15,16,17,18,19,20,21,22,23,24},
]
\addplot+[ycomb,bluePlots,mark=*,mark options={scale=0.5, fill=bluePlots},] 
table {%
1     2.327847
2     1.359034
3     1.605471
4     2.280212
5     1.228364
6     0.163737
7     0.771787
8     2.103309
9     2.638321
10    2.473109
11    1.262122
12    1.936341
13    2.105984
14    1.493746
15    0.951190
16    0.781059
17    0.877848
18   -0.427474
19   -0.231807
20   -0.174024
21    0.980250
22    0.352188
23    0.565517
24   -0.208062
};
%plot coordinates {
%	(0,3) 
%	(1,2) 
%	(2,4) 
%	(3,1) 
%	(4,1)
%	(5,1)
%	(6,1)
%	(7,1)
%	(8,1)
%	(9,1)
%	(10,1)
%	(11,1)
%	(12,1)
%	(13,1)
%	(14,1)
%	(15,1)
%	(16,1)
%	(17,1)
%	(18,1)
%	(19,1)
%	(20,1)
%	(21,1)
%	(22,1)
%	(23,1)
%	(24,1)};
	
\addplot [color=black,line width=0.15mm,dashed,forget plot]
table[row sep=crcr]{%
	0 1.6448536269514722\\
	25 1.6448536269514722\\
};	

\addplot [color=black,line width=0.15mm,dashed,forget plot]
table[row sep=crcr]{%
	0 -1.6448536269514722\\
	25 -1.6448536269514722\\
};	

\addplot [color=black,line width=0.65mm,thick,forget plot]
table[row sep=crcr]{%
	0 0\\
	25 0\\
};	
	
\end{axis} 
\end{tikzpicture}
		\vspace{-0.5cm}
		\caption{ DM test results when comparing $M_{\mr{Single}}$ and $M_{\mr{{Dual}}}$. Values above the top dashed line represent cases where, with a 95 \% confidence level, $M_{\mr{{Dual}}}$ is significantly better. Similarly, values below the lower dashed line accept, with a 95 \% confidence level, that $M_{\mr{{Dual}}}$ is significantly worse.}
		\label{fig:ex3DM}
	\end{center}
\end{figure}
As we can see from the plot, the forecast improvements of the dual-market forecaster are statistically significant in 7 of the 24 day-ahead prices. In addition, the single-market forecaster is not significantly better in any of the remaining 17 day-ahead prices. Therefore, as $M_{\mr{{Dual}}}$ is approximately better for a third of the day-ahead prices and not worse for the remaining two-thirds, we can conclude that the dual-market forecaster is a statistically significant better forecaster.

Finally, we also perform the DM test on the full loss differential considering serial correlation. Once again, the obtained metrics agree with the results obtained for the individual 24 hours: with a $p$-value of \expnumber{9.5}{-03}, the test results confirm the statistical significance of the difference in predictive accuracy when using the dual-market forecaster.

\subsection{Analysis and Discussion}
\label{sec:65}
To understand and explain the obtained results, we have to note that, as introduced in Section \ref{sec:motivation}, market integration across European electricity markets has been increasing over the years due to EU regulations. This highly nonlinear and complex effect dramatically modifies the dynamics of electricity prices and is behind the obtained improvements of our models. In particular, our forecasters use this effect to outperform alternative techniques that have traditionally ignored it: the first forecaster proposed, which models market integration in the input space, obtains statistically significant improvements w.r.t.~to model counterparts that disregard market integration. The second proposed forecaster, which goes one step further by modeling market integration in the output space, is shown to be crucial to obtain further significant improvements. For our case study, this translates to the following conclusions:
\begin{enumerate}
\item Using features from the French market significantly enhances the predictive accuracy of a model forecasting Belgian prices. The results are statistically significant and independent of whether Belgian features are considered or not.
\item A dual-market forecaster simultaneously predicting prices in France and Belgium can improve the predictive accuracy. In particular, by solving two related tasks, it is able to learn more useful features, to better generalize the price dynamics, and to obtain improvements that are statistically significant.
\item The proposed feature selection algorithm is able to perform a correct assessment of the importance of features.
\end{enumerate}

In addition, it is interesting to see how explanatory variables from the EPEX-Belgium, e.g.~load and generation, have almost no influence in the day-ahead prices. In fact, from the obtained results, it is surprising to observe how French factors play a larger role in Belgian prices than the local Belgian features.

As a final discussion, it is necessary to indicate why, while being neighboring countries of Belgium, The Netherlands and Germany and their respective markets have not been considered in the study. The reason for not considering The Netherlands is the fact that the amount of available online data is smaller than in France and Belgium, and thus, training the DNNs can be harder. In the case of Germany, the reason for not considering it is that, at the moment, there is not a direct interconnection of the electrical grid between Belgium and Germany.

\subsection{Practical Applications}
\label{sec:66}
As a last remark, it is important to point out the different practical applications that these results have. Particularly, there are two main obvious applications where this research can be highly beneficial. The first and most important application is its usage by utility companies to increase their economic profit. More specifically, a 1\% improvement in the MAPE of the forecasting accuracy results
in about 0.1\%-0.35\% cost reduction \cite{Zareipour2010}. For a medium-size utility company with a peak load of 5 GW, this translates to saving approximately \$1.5 million per year \cite{Hong2015,Uniejewski2016}. 

In addition, improvements in forecasting accuracy are key to have a stable electrical grid. Particularly, as the integration of renewable energy sources increases, so do the imbalances in the electrical grid due to mismatches between generation and consumption. To tackle this issue demand response methods \cite{Wang2017,Nolan2015,Wang2015} have been traditionally applied. By accurate forecasting electricity prices it is also possible to improve the situation. In particular, prices are usually low (high) when generation is larger (lower) than consumption. Therefore, given the right forecasts, market agents have economic incentives to buy (sell) energy when prices are low (high), and in turn, to reduce the grid imbalances. Therefore, using accurate price forecasting, market agents can steered and motivated so that grid imbalances are reduced.

\section{Conclusions}
\label{sec:conc}
We have analyzed how market integration can be used to enhance the predictive accuracy of day-ahead price forecasting in electricity markets. In particular, we have proposed a first model that, by considering features from connected markets, improves the predictive performance. In addition, we have proposed a dual-market forecaster that, by multitasking and due to market integration, can further improve the predictive accuracy. As a case study, we have considered the electricity markets in Belgium and France. Then, we have showed how, considering market integration, the proposed forecasters lead to improvements that are statistically significant. Additionally, we have proposed a novel feature selection algorithm and using the same case study, we have shown how the algorithm correctly assesses feature importance.

In view of these results, it is clear that market integration can play a large role in electricity prices. In particular, the influence of neighboring markets seems to be important enough to build statistically significant differences in terms of forecasting accuracy. As a consequence, as the EU has implemented regulations to form an integrated EU market but there is still little insight in the outcome of such regulations, these results are important in terms of policy making. In particular, the fact that market integration largely modifies the price dynamics between Belgium and France is an indicator that the regulations that were put in place are working. As a result, using the proposed methodology, policy makers can benefit from a general tool to evaluate the market integration regulations in other EU regions.	

In addition, these results are also of high importance in terms of grid stability and economic profit of market agents. In particular, as the knowledge of the dynamics of electricity prices increases, the grid operator might be able to better prevent some of the grid imbalances characterized by large price peaks. The increased knowledge is also economically beneficial for market agents: a 1 \% improvement in MAPE accuracy translates to savings of \$1.5 million per year for a medium-size utility company.

As a first step to help policy markets, in future work the performed experiments will be expanded to the other European markets.

\section*{Acknowledgment}
This research has received funding from the European Union’s Horizon 2020 research and innovation program under the Marie Skłodowska-Curie grant agreement No 675318 (INCITE).

\section*{Copyright Information}

\noindent \copyright~2017. This manuscript version is made available under the CC-BY-NC-ND 4.0 license \url{http://creativecommons.org/licenses/by-nc-nd/4.0/}.
\vspace{0.25em}

\noindent \hspace{-0.4em}\doclicenseImage[imagewidth=6em]

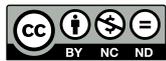
\bibliography{bibtex/bibtex}

\begin{thebibliography}{10}
\expandafter\ifx\csname url\endcsname\relax
  \def\url#1{\texttt{#1}}\fi
\expandafter\ifx\csname urlprefix\endcsname\relax\def\urlprefix{URL }\fi
\expandafter\ifx\csname href\endcsname\relax
  \def\href#1#2{#2} \def\path#1{#1}\fi

\bibitem{Weron2014}
R.~Weron, {Electricity price forecasting: A review of the state-of-the-art with
  a look into the future}, International Journal of Forecasting 30~(4) (2014)
  1030--1081.
\newblock \href {http://dx.doi.org/10.1016/j.ijforecast.2014.08.008}
  {\path{doi:10.1016/j.ijforecast.2014.08.008}}.

\bibitem{Jamasb2005}
T.~Jamasb, M.~Pollitt, Electricity market reform in the {European} union:
  review of progress toward liberalization \& integration, The Energy Journal
  26 (2005) 11--41.
\newblock \href {http://dx.doi.org/10.5547/issn0195-6574-ej-vol26-nosi-2}
  {\path{doi:10.5547/issn0195-6574-ej-vol26-nosi-2}}.

\bibitem{Weron2008}
R.~Weron, A.~Misiorek, Forecasting spot electricity prices: A comparison of
  parametric and semiparametric time series models, International Journal of
  Forecasting 24~(4) (2008) 744--763.
\newblock \href {http://dx.doi.org/10.1016/j.ijforecast.2008.08.004}
  {\path{doi:10.1016/j.ijforecast.2008.08.004}}.

\bibitem{CrespoCuaresma2004}
J.~Crespo~Cuaresma, J.~Hlouskova, S.~Kossmeier, M.~Obersteiner, Forecasting
  electricity spot-prices using linear univariate time-series models, Applied
  Energy 77~(1) (2004) 87--106.
\newblock \href {http://dx.doi.org/10.1016/S0306-2619(03)00096-5}
  {\path{doi:10.1016/S0306-2619(03)00096-5}}.

\bibitem{Yang2017}
Z.~Yang, L.~Ce, L.~Lian, Electricity price forecasting by a hybrid model,
  combining wavelet transform, {ARMA} and kernel-based extreme learning machine
  methods, Applied Energy 190 (2017) 291--305.
\newblock \href {http://dx.doi.org/10.1016/j.apenergy.2016.12.130}
  {\path{doi:10.1016/j.apenergy.2016.12.130}}.

\bibitem{Nogales2002}
F.~J. Nogales, J.~Contreras, A.~J. Conejo, R.~Esp{\'{i}}nola, Forecasting
  next-day electricity prices by time series models, IEEE Transactions on Power
  Systems 17~(2) (2002) 342--348.
\newblock \href {http://dx.doi.org/10.1109/MPER.2002.4312063}
  {\path{doi:10.1109/MPER.2002.4312063}}.

\bibitem{Cruz2011}
A.~Cruz, A.~Mu{\~{n}}oz, J.~Zamora, R.~Esp{\'{i}}nola, The effect of wind
  generation and weekday on {S}panish electricity spot price forecasting,
  Electric Power Systems Research 81~(10) (2011) 1924--1935.
\newblock \href {http://dx.doi.org/10.1016/j.epsr.2011.06.002}
  {\path{doi:10.1016/j.epsr.2011.06.002}}.

\bibitem{Misiorek2006}
A.~Misiorek, S.~Trueck, R.~Weron, Point and interval forecasting of spot
  electricity prices: Linear vs. non-linear time series models, Studies in
  Nonlinear Dynamics {\&} Econometrics 10~(3) (2006) 1--36.
\newblock \href {http://dx.doi.org/10.2202/1558-3708.1362}
  {\path{doi:10.2202/1558-3708.1362}}.

\bibitem{Diongue2009}
A.~K. Diongue, D.~Guégan, B.~Vignal, Forecasting electricity spot market
  prices with a k-factor {GIGARCH} process, Applied Energy 86~(4) (2009)
  505--510.
\newblock \href {http://dx.doi.org/10.1016/j.apenergy.2008.07.005}
  {\path{doi:10.1016/j.apenergy.2008.07.005}}.

\bibitem{Conejo2005}
A.~Conejo, M.~Plazas, R.~Espinola, A.~Molina, Day-ahead electricity price
  forecasting using the wavelet transform and {ARIMA} models, IEEE Transactions
  on Power Systems 20~(2) (2005) 1035--1042.
\newblock \href {http://dx.doi.org/10.1109/TPWRS.2005.846054}
  {\path{doi:10.1109/TPWRS.2005.846054}}.

\bibitem{Tan2010}
Z.~Tan, J.~Zhang, J.~Wang, J.~Xu, Day-ahead electricity price forecasting using
  wavelet transform combined with {ARIMA} and {GARCH} models, Applied Energy
  87~(11) (2010) 3606--3610.
\newblock \href {http://dx.doi.org/10.1016/j.apenergy.2010.05.012}
  {\path{doi:10.1016/j.apenergy.2010.05.012}}.

\bibitem{Amjady2006a}
N.~Amjady, M.~Hemmati, Energy price forecasting - problems and proposals for
  such predictions, {IEEE} Power and Energy Magazine 4~(2) (2006) 20--29.
\newblock \href {http://dx.doi.org/10.1109/MPAE.2006.1597990}
  {\path{doi:10.1109/MPAE.2006.1597990}}.

\bibitem{Szkuta1999}
B.~Szkuta, L.~Sanabria, T.~Dillon, Electricity price short-term forecasting
  using artificial neural networks, IEEE Transactions on Power Systems 14~(3)
  (1999) 851--857.
\newblock \href {http://dx.doi.org/10.1109/59.780895}
  {\path{doi:10.1109/59.780895}}.

\bibitem{Catalao2007}
J.~P.~S. Catal{\~a}o, S.~J. P.~S. Mariano, V.~M.~F. Mendes, L.~A. F.~M.
  Ferreira, Short-term electricity prices forecasting in a competitive market:
  A neural network approach, Electric Power Systems Research 77~(10) (2007)
  1297--1304.
\newblock \href {http://dx.doi.org/10.1016/j.epsr.2006.09.022}
  {\path{doi:10.1016/j.epsr.2006.09.022}}.

\bibitem{Xiao2017}
L.~Xiao, W.~Shao, M.~Yu, J.~Ma, C.~Jin, Research and application of a hybrid
  wavelet neural network model with the improved cuckoo search algorithm for
  electrical power system forecasting, Applied Energy 198 (2017) 203--222.
\newblock \href {http://dx.doi.org/10.1016/j.apenergy.2017.04.039}
  {\path{doi:10.1016/j.apenergy.2017.04.039}}.

\bibitem{Wang2017a}
D.~Wang, H.~Luo, O.~Grunder, Y.~Lin, H.~Guo, Multi-step ahead electricity price
  forecasting using a hybrid model based on two-layer decomposition technique
  and {BP} neural network optimized by firefly algorithm, Applied Energy 190
  (2017) 390--407.
\newblock \href {http://dx.doi.org/10.1016/j.apenergy.2016.12.134}
  {\path{doi:10.1016/j.apenergy.2016.12.134}}.

\bibitem{Fan2007}
S.~Fan, C.~Mao, L.~Chen, Next-day electricity-price forecasting using a hybrid
  network, IET Generation, Transmission \& Distribution 1~(1) (2007) 176--182.
\newblock \href {http://dx.doi.org/10.1049/iet-gtd:20060006}
  {\path{doi:10.1049/iet-gtd:20060006}}.

\bibitem{Lin2010}
W.-M. Lin, H.-J. Gow, M.-T. Tsai, An enhanced radial basis function network for
  short-term electricity price forecasting, Applied Energy 87~(10) (2010)
  3226--3234.
\newblock \href {http://dx.doi.org/10.1016/j.apenergy.2010.04.006}
  {\path{doi:10.1016/j.apenergy.2010.04.006}}.

\bibitem{Amjady2006}
N.~Amjady, Day-ahead price forecasting of electricity markets by a new fuzzy
  neural network, IEEE Transactions on Power Systems 21~(2) (2006) 887--896.
\newblock \href {http://dx.doi.org/10.1109/tpwrs.2006.873409}
  {\path{doi:10.1109/tpwrs.2006.873409}}.

\bibitem{Lago2017}
J.~Lago, F.~{De Ridder}, B.~{De Schutter}, Forecasting spot electricity prices:
  deep learning approaches and empirical comparison of traditional algorithms,
  Applied Energy (Submitted).

\bibitem{Meeus2008}
L.~Meeus, R.~Belmans, Electricity market integration in {E}urope, in:
  Proceedings of the 16th Power Systems Computation Conference, 2008.

\bibitem{Bunn2010}
D.~W. Bunn, A.~Gianfreda, Integration and shock transmissions across {European}
  electricity forward markets, Energy Economics 32~(2) (2010) 278--291.
\newblock \href {http://dx.doi.org/10.1016/j.eneco.2009.09.005}
  {\path{doi:10.1016/j.eneco.2009.09.005}}.

\bibitem{DeMenezes2016}
L.~M. de~Menezes, M.~A. Houllier, Reassessing the integration of {European}
  electricity markets: A fractional cointegration analysis, Energy Economics 53
  (2016) 132--150.
\newblock \href {http://dx.doi.org/10.1016/j.eneco.2014.10.021}
  {\path{doi:10.1016/j.eneco.2014.10.021}}.

\bibitem{Zachmann2008}
G.~Zachmann, Electricity wholesale market prices in {Europe}: {Convergence}?,
  Energy Economics 30~(4) (2008) 1659--1671.
\newblock \href {http://dx.doi.org/10.1016/j.eneco.2007.07.002}
  {\path{doi:10.1016/j.eneco.2007.07.002}}.

\bibitem{Lindstroem2012}
E.~Lindstr{\"o}m, F.~Regland, Modeling extreme dependence between {E}uropean
  electricity markets, Energy Economics 34~(4) (2012) 899--904.
\newblock \href {http://dx.doi.org/10.1016/j.eneco.2012.04.006}
  {\path{doi:10.1016/j.eneco.2012.04.006}}.

\bibitem{Ziel2015}
F.~Ziel, R.~Steinert, S.~Husmann, Forecasting day ahead electricity spot
  prices: The impact of the {EXAA} to other {European} electricity markets,
  Energy Economics 51 (2015) 430--444.
\newblock \href {http://dx.doi.org/10.1016/j.eneco.2015.08.005}
  {\path{doi:10.1016/j.eneco.2015.08.005}}.

\bibitem{Panapakidis2016}
I.~P. Panapakidis, A.~S. Dagoumas, Day-ahead electricity price forecasting via
  the application of artificial neural network based models, Applied Energy 172
  (2016) 132--151.
\newblock \href {http://dx.doi.org/10.1016/j.apenergy.2016.03.089}
  {\path{doi:10.1016/j.apenergy.2016.03.089}}.

\bibitem{Guyon2003}
I.~Guyon, A.~Elisseeff, An introduction to variable and feature selection,
  Journal of Machine Learning Research 3 (2003) 1157--1182.

\bibitem{Carta2015}
J.~A. Carta, P.~Cabrera, J.~M. Matías, F.~Castellano, Comparison of feature
  selection methods using {ANNs} in {MCP}-wind speed methods. a case study,
  Applied Energy 158 (2015) 490--507.
\newblock \href {http://dx.doi.org/10.1016/j.apenergy.2015.08.102}
  {\path{doi:10.1016/j.apenergy.2015.08.102}}.

\bibitem{Goodfellow2016}
I.~Goodfellow, Y.~Bengio, A.~Courville, Deep Learning, MIT Press, 2016,
  \url{http://www.deeplearningbook.org/}.

\bibitem{Stevenson2001}
M.~Stevenson, Filtering and forecasting spot electricity prices in the
  increasingly deregulated australian electricity market, in: QFRC Research
  Paper Series, no.~63, Quantitative Finance Research Centre, University of
  Technology, Sydney, 2001,
  \url{http://www.qfrc.uts.edu.au/research/research_papers/rp63.pdf}.

\bibitem{Rodriguez2004}
C.~P. Rodriguez, G.~J. Anders, Energy price forecasting in the {Ontario}
  competitive power system market, IEEE Transactions on Power Systems 19~(1)
  (2004) 366--374.
\newblock \href {http://dx.doi.org/10.1109/TPWRS.2003.821470}
  {\path{doi:10.1109/TPWRS.2003.821470}}.

\bibitem{Hong2012}
Y.~Hong, C.~Wu, Day-ahead electricity price forecasting using a hybrid
  principal component analysis network, Energies 5~(11) (2012) 4711--4725.
\newblock \href {http://dx.doi.org/10.3390/en5114711}
  {\path{doi:10.3390/en5114711}}.

\bibitem{Amjady2009}
N.~Amjady, F.~Keynia, Day-ahead price forecasting of electricity markets by
  mutual information technique and cascaded neuro-evolutionary algorithm, IEEE
  Transactions on Power Systems 24~(1) (2009) 306--318.
\newblock \href {http://dx.doi.org/10.1109/tpwrs.2008.2006997}
  {\path{doi:10.1109/tpwrs.2008.2006997}}.

\bibitem{Amjady2010}
N.~Amjady, A.~Daraeepour, F.~Keynia, Day-ahead electricity price forecasting by
  modified relief algorithm and hybrid neural network, IET Generation,
  Transmission \& Distribution 4~(3) (2010) 432--444.
\newblock \href {http://dx.doi.org/10.1049/iet-gtd.2009.0297}
  {\path{doi:10.1049/iet-gtd.2009.0297}}.

\bibitem{Keles2016}
D.~Keles, J.~Scelle, F.~Paraschiv, W.~Fichtner, Extended forecast methods for
  day-ahead electricity spot prices applying artificial neural networks,
  Applied Energy 162 (2016) 218--230.
\newblock \href {http://dx.doi.org/10.1016/j.apenergy.2015.09.087}
  {\path{doi:10.1016/j.apenergy.2015.09.087}}.

\bibitem{Ghasemi2016}
A.~Ghasemi, H.~Shayeghi, M.~Moradzadeh, M.~Nooshyar, A novel hybrid algorithm
  for electricity price and load forecasting in smart grids with demand-side
  management, Applied Energy 177 (2016) 40--59.
\newblock \href {http://dx.doi.org/10.1016/j.apenergy.2016.05.083}
  {\path{doi:10.1016/j.apenergy.2016.05.083}}.

\bibitem{Abedinia2017}
O.~Abedinia, N.~Amjady, H.~Zareipour, A new feature selection technique for
  load and price forecast of electrical power systems, IEEE Transactions on
  Power Systems 32~(1) (2017) 62--74.
\newblock \href {http://dx.doi.org/10.1109/TPWRS.2016.2556620}
  {\path{doi:10.1109/TPWRS.2016.2556620}}.

\bibitem{Ruder2016}
S.~Ruder, An overview of gradient descent optimization algorithms, arXiv eprint
  (2016).
\newblock \href {http://arxiv.org/abs/1609.04747} {\path{arXiv:1609.04747}}.

\bibitem{Shafie2011}
M.~Shafie-Khah, M.~P. Moghaddam, M.~Sheikh-El-Eslami, Price forecasting of
  day-ahead electricity markets using a hybrid forecast method, Energy
  Conversion and Management 52~(5) (2011) 2165--2169.
\newblock \href {http://dx.doi.org/10.1016/j.enconman.2010.10.047}
  {\path{doi:10.1016/j.enconman.2010.10.047}}.

\bibitem{Jones1998}
D.~R. Jones, M.~Schonlau, W.~J. Welch, Efficient global optimization of
  expensive black-box functions, Journal of Global Optimization 13~(4) (1998)
  455--492.
\newblock \href {http://dx.doi.org/10.1023/A:1008306431147}
  {\path{doi:10.1023/A:1008306431147}}.

\bibitem{Bergstra2011}
J.~Bergstra, R.~Bardenet, Y.~Bengio, B.~K{\'{e}}gl, Algorithms for
  hyper-parameter optimization, in: Advances in Neural Information Processing
  Systems, 2011, pp. 2546--2554,
  \url{http://papers.nips.cc/paper/4443-algorithms-for-hyper-parameter-optimization}.

\bibitem{Hutter2011}
F.~Hutter, H.~H. Hoos, K.~Leyton-Brown, Sequential model-based optimization for
  general algorithm configuration, in: International Conference on Learning and
  Intelligent Optimization, Springer, 2011, pp. 507--523.
\newblock \href {http://dx.doi.org/10.1007/978-3-642-25566-3_40}
  {\path{doi:10.1007/978-3-642-25566-3_40}}.

\bibitem{Hutter2014}
F.~Hutter, H.~Hoos, K.~Leyton-Brown, An efficient approach for assessing
  hyperparameter importance, in: Proceedings of the 31st International
  Conference on International Conference on Machine Learning, Vol.~32 of
  ICML'14, 2014, pp. 754--762,
  \url{http://proceedings.mlr.press/v32/hutter14.pdf}.

\bibitem{Makridakis1993}
S.~Makridakis, Accuracy measures: theoretical and practical concerns,
  International Journal of Forecasting 9~(4) (1993) 527--529.
\newblock \href {http://dx.doi.org/10.1016/0169-2070(93)90079-3}
  {\path{doi:10.1016/0169-2070(93)90079-3}}.

\bibitem{Diebold1995}
F.~X. Diebold, R.~S. Mariano, Comparing predictive accuracy, Journal of
  Business {\&} Economic Statistics 13~(3) (1995) 253--263.
\newblock \href {http://dx.doi.org/10.1080/07350015.1995.10524599}
  {\path{doi:10.1080/07350015.1995.10524599}}.

\bibitem{LeCun1998}
Y.~LeCun, L.~Bottou, G.~B. Orr, K.-R. Müller, Efficient {BackProp}, in: G.~B.
  Orr, K.-R. Müller (Eds.), Neural {Networks}: {Tricks} of the {Trade}, no.
  1524 in Lecture {Notes} in {Computer} {Science}, Springer Berlin Heidelberg,
  1998, pp. 9--50.
\newblock \href {http://dx.doi.org/10.1007/3-540-49430-8_2}
  {\path{doi:10.1007/3-540-49430-8_2}}.

\bibitem{RTE}
RTE, Grid data, \url{https://data.rte-france.com/}. Accessed on 15.05.2017.

\bibitem{Elia}
Elia, Grid data, \url{http://www.elia.be/en/grid-data/dashboard}. Accessed on
  15.05.2017.

\bibitem{entsoe}
{ENTSO-E} transparency platform, \url{https://transparency.entsoe.eu/}.
  Accessed on 15.05.2017.

\bibitem{Nair2010}
V.~Nair, G.~E. Hinton, Rectified linear units improve restricted boltzmann
  machines, in: Proceedings of the 27th international Conference on Machine
  Learning (ICML), 2010, pp. 807--814,
  \url{http://icml2010.haifa.il.ibm.com/papers/432.pdf}.

\bibitem{Glorot2010}
X.~Glorot, Y.~Bengio, Understanding the difficulty of training deep feedforward
  neural networks, in: Proceedings of the International Conference on
  Artificial Intelligence and Statistics ({AISTATS}’10). Society for
  Artificial Intelligence and Statistics, 2010, pp. 249--256.

\bibitem{Kingma2014}
D.~P. Kingma, J.~Ba, Adam: A method for stochastic optimization, arXiv eprint
  (2014).
\newblock \href {http://arxiv.org/abs/1412.6980} {\path{arXiv:1412.6980}}.

\bibitem{Yao2007}
Y.~Yao, L.~Rosasco, A.~Caponnetto, On early stopping in gradient descent
  learning, Constructive Approximation 26~(2) (2007) 289--315.
\newblock \href {http://dx.doi.org/10.1007/s00365-006-0663-2}
  {\path{doi:10.1007/s00365-006-0663-2}}.

\bibitem{Yosinski2014}
J.~Yosinski, J.~Clune, Y.~Bengio, H.~Lipson, How transferable are features in
  deep neural networks?, in: Z.~Ghahramani, M.~Welling, C.~Cortes, N.~D.
  Lawrence, K.~Q. Weinberger (Eds.), Advances in {Neural} {Information}
  {Processing} {Systems} 27, Curran Associates, Inc., 2014, pp. 3320--3328,
  \url{https://papers.nips.cc/paper/5347-how-transferable-are-features-in-deep-neural-networks}.

\bibitem{Jaderberg2016}
M.~Jaderberg, V.~Mnih, W.~M. Czarnecki, T.~Schaul, J.~Z. Leibo, D.~Silver,
  K.~Kavukcuoglu, Reinforcement learning with unsupervised auxiliary tasks,
  arXiv eprint (2016).
\newblock \href {http://arxiv.org/abs/1611.05397} {\path{arXiv:1611.05397}}.

\bibitem{Li2016}
X.~Li, L.~Zhao, L.~Wei, M.-H. Yang, F.~Wu, Y.~Zhuang, H.~Ling, J.~Wang,
  {DeepSaliency}: {Multi}-{Task} {Deep} {Neural} {Network} model for salient
  object detection, IEEE Transactions on Image Processing 25~(8) (2016)
  3919--3930.
\newblock \href {http://dx.doi.org/10.1109/TIP.2016.2579306}
  {\path{doi:10.1109/TIP.2016.2579306}}.

\bibitem{Bergstra2013}
J.~Bergstra, D.~Yamins, D.~D. Cox, Making a science of model search:
  Hyperparameter optimization in hundreds of dimensions for vision
  architectures, in: Proceedings of the 30th International Conference on
  Machine Learning, 2013, pp. 115--123,
  \url{http://proceedings.mlr.press/v28/bergstra13.pdf}.

\bibitem{Zareipour2010}
H.~Zareipour, C.~A. Canizares, K.~Bhattacharya, Economic impact of electricity
  market price forecasting errors: A demand-side analysis, {IEEE} Transactions
  on Power Systems 25~(1) (2010) 254--262.
\newblock \href {http://dx.doi.org/10.1109/TPWRS.2009.2030380}
  {\path{doi:10.1109/TPWRS.2009.2030380}}.

\bibitem{Hong2015}
T.~Hong, Crystal ball lessons in predictive analytics, EnergyBiz 12~(2) (2015)
  35--37.

\bibitem{Uniejewski2016}
B.~Uniejewski, J.~Nowotarski, R.~Weron, Automated variable selection and
  shrinkage for day-ahead electricity price forecasting, Energies 9~(8) (2016)
  621.
\newblock \href {http://dx.doi.org/10.3390/en9080621}
  {\path{doi:10.3390/en9080621}}.

\bibitem{Wang2017}
J.~Wang, H.~Zhong, Z.~Ma, Q.~Xia, C.~Kang, Review and prospect of integrated
  demand response in the multi-energy system, Applied Energy 202 (2017)
  772--782.
\newblock \href {http://dx.doi.org/10.1016/j.apenergy.2017.05.150}
  {\path{doi:10.1016/j.apenergy.2017.05.150}}.

\bibitem{Nolan2015}
S.~Nolan, M.~O’Malley, Challenges and barriers to demand response deployment
  and evaluation, Applied Energy 152 (2015) 1--10.
\newblock \href {http://dx.doi.org/10.1016/j.apenergy.2015.04.083}
  {\path{doi:10.1016/j.apenergy.2015.04.083}}.

\bibitem{Wang2015}
Q.~Wang, C.~Zhang, Y.~Ding, G.~Xydis, J.~Wang, J.~Østergaard, Review of
  real-time electricity markets for integrating distributed energy resources
  and demand response, Applied Energy 138 (2015) 695--706.
\newblock \href {http://dx.doi.org/10.1016/j.apenergy.2014.10.048}
  {\path{doi:10.1016/j.apenergy.2014.10.048}}.

\end{thebibliography}

\end{document}